\documentclass[aps,prb,onecolumn,superscriptaddress,longbibliography]{revtex4-2}

\usepackage{graphicx}
\usepackage{amsmath,amssymb,bm}
\usepackage{xcolor}
\usepackage{comment}
\PassOptionsToPackage{hyphens}{url}
\usepackage{hyperref}

\makeatletter
\g@addto@macro\UrlBreaks{\do\a\do\b\do\c\do\d\do\e\do\f\do\g\do\h\do\i
  \do\j\do\k\do\l\do\m\do\n\do\o\do\p\do\q\do\r\do\s\do\t\do\u
  \do\v\do\w\do\x\do\y\do\z\do\0\do\1\do\2\do\3\do\4\do\5\do\6
  \do\7\do\8\do\9}
\makeatother
\graphicspath{{figs/}}

\newcommand{\avh}[1]{\langle #1 \rangle_{h}}
\newcommand{\avmu}[1]{\langle #1 \rangle_{\mu_1}}

\newcommand{\Tr}{\operatorname{Tr}}

\newcommand{\sign}{\operatorname{sign}}

\begin{document}
\title{Quenched complexity of marginal states in the Sherrington--Kirkpatrick spin glass}

\author{Tiziana de Chirico}
\email{tiziana.dechirico@uniroma1.it}
\affiliation{Dipartimento di Fisica, Sapienza Universit\`a di Roma, Piazzale Aldo Moro 5, I-00185 Rome, Italy}
\affiliation{Institute of Nanotechnology, CNR-NANOTEC, via Monteroni, Lecce 73100, Italy}

\author{Luca Leuzzi}
\email{luca.leuzzi@cnr.it}
\affiliation{Institute of Nanotechnology, CNR-NANOTEC, Soft and Living Matter Laboratory, Piazzale Aldo Moro 5, I-00185 Rome, Italy}
\affiliation{Dipartimento di Fisica, Sapienza Universit\`a di Roma, Piazzale Aldo Moro 5, I-00185 Rome, Italy}
\affiliation{
INFN, Sezione di Roma I, P.le Aldo Moro 5,  I-00185, Rome, Italy}

\date{\today}

\begin{abstract}
The number of metastable states of a random corrugated landscape controls what dynamics, inference and learning can reach. In the Sherrington-Kirkpatrick model these states are marginal, and counting them requires breaking the BRST symmetry. So far only the annealed complexity was known, predicting states below equilibrium. We compute the quenched complexity function with a two-group Ansatz and one-step replica symmetry breaking. It branches continuously off the annealed curve at its instability point and vanishes near the Parisi free energy, while the soft mode turns toward orthogonality to the magnetization.
\end{abstract}

\maketitle

The statistical geometry of high-dimensional random landscapes has become a
common language for disordered systems, inference and machine learning: how
many stationary points a random cost function possesses, and how stable they
are, controls which configurations an algorithm can actually reach
\cite{Ros19,Ros23}, it sets the reach of the message-passing algorithms that
approach the Parisi ground-state energy of the Sherrington--Kirkpatrick model
itself \cite{Montanari19,ElAlaoui21}, and, in
neural networks, it selects wide, high-local-entropy regions rather than the
deepest isolated minima \cite{Baldassi15}. In all these settings, the
organizing quantity is called the
complexity.
Mean-field glassy systems possess a 
landscape populated by an exponential number of metastable states $\mathcal{N}(f)$ at each (free) energy density $f$, whose
counting is quantified by the complexity, or configurational entropy,
$\Sigma(f) = N^{-1}\ln \mathcal{N}(f)$
\cite{Thouless77,Bray80,Crisanti92,Monasson95,Cavagna98b,Leuzzi05}.
In models with a one-step replica symmetry broken (1RSB) equilibrium phase, such as the
spherical $p$-spin model \cite{Crisanti92}, metastable states at low free energy are genuine
minima of the Thouless--Anderson--Palmer (TAP) free energy
\cite{Thouless77,Crisanti95,Crisanti03b}, and their complexity can be computed within a
 Becchi--Rouet--Stora--Tyutin (BRST) symmetric
\cite{Becchi75,ZinnJustin02} saddle point of the counting action
\cite{Annibale03,Annibale04}.
When the states are, instead, mutually correlated, the counting 
acquires a nontrivial replica structure.
For the mixed spherical models, whose
equilibrium phase diagram already hosts 1RSB, full replica symmetry broken (FRSB), mixed discrete and FRSB phases and glass-to-glass
transitions~\cite{Crisanti04c,Crisanti06,Crisanti07b}, the complexity of the
thermodynamic states was obtained long ago as a Legendre transform of the
replicated free energy evaluated in the 1RSB and 1-FRSB
Ansatz~\cite{Crisanti06}; more recently, a hierarchical (1RSB and FRSB)
Ansatz has been formulated for the Kac--Rice count of stationary points, with
replica symmetry breaking in the space of the counted points
themselves~\cite{KentDobias23}, correctly reducing to the Parisi ground state
at low temperature. The criterion for the failure of the annealed count, a
finite fraction of stationary points with nontrivial mutual correlations, has
been characterized in general~\cite{KentDobias23b}.

The complex high dimensional landscape constrains, as well, off-equilibrium dynamics, a link that is exact only for the pure spherical $p$-spin model \cite{Cugliandolo93}. In mixed spherical models, long-time integration of the dynamical mean-field equations \cite{Lang25,Lang26} reveals strong, not weak \cite{Folena20,Folena23}, ergodicity breaking \cite{Citro25}. Post-quench aging states fall into phases with one, two or a continuum of effective temperatures, which decide whether gradient flow reaches the algorithmic energy bound \cite{Lang26b}. They also require a generalized Cugliandolo--Kurchan Ansatz, whose three-time composition law is in general non-analytic \cite{Lang26c}.


The Sherrington Kirkpatrick (SK) model \cite{Sherrington75,Mezard87,Parisi79,Parisi83} turns out to be radically
different from these systems: below the spin-glass transition temperature $T_c$
 the overwhelming majority of its TAP states
are {\em marginal} at all free energies, that is, each possesses a soft mode of the
free-energy Hessian \cite{Bray81,Aspelmeier04,Mueller06}. Numerical and analytical evidence further suggests that TAP minima are not isolated but, at finite $N$, come in as pairs of minima and saddle of rank one
\cite{Cavagna2004,Aspelmeier04}.
Marginal states escape the BRST-symmetric counting: their description requires
breaking this fermionic symmetry \cite{Parisi95c,Crisanti03c,Parisi04b,Mueller06}.
The two-group (2G) replica Ansatz, originally introduced by Bray and Moore \cite{Bray78} and adapted to this context by Parisi and Potters \cite{Parisi95c}, allows to derive the BRST-symmetry-broken complexity as the Legendre transform of a generalized two-group  free energy functional. Physically, the two groups of
replicas represent a minimum and a rank-one saddle coalescing into a single marginal state in some limit. The emerging order parameters encode the correlations
between the local magnetization and the soft mode characterizing each marginal state \cite{Mueller06}. 

Despite this 45 years long history, as far as we know, the two-group complexity of the SK model has been computed only in the annealed approximation
\cite{Bray80,Parisi95c,Crisanti03c,Mueller06}, i.e., averaging the number of states, rather
than its logarithm, over the disorder. 

The Bray--Moore annealed complexity $\Sigma_{\rm ann}(f)$ is exact only for $f^*<f<f_{\rm max}$ \cite{Bray81,Crisanti03c}; below $f^*$ a replicon eigenvalue of the two-group Hessian turns negative. Moreover, $\Sigma_{\rm ann}$ vanishes at $f_0^{\rm ann}<f_0^{\rm eq}$, which is unphysical, since no state can lie below equilibrium: at $T=0$, $e_0^{\rm ann}=-0.79067$ \cite{Bray80,Crisanti03c} versus $e_0^{\rm eq}=-0.76317$ \cite{Parisi79,Crisanti02b,SchmidtOppermann08}.  A recent Kac--Rice reformulation in a four-index superspace handles the modulus of the Hessian determinant through a minimal spontaneous symmetry breaking, and gives both total and marginal annealed complexities \cite{KentDobias26}. Rigorous annealed results, with a Monasson-like Legendre structure, exist for the generalized TAP free energy \cite{Chen23,Chen21,Boursier26}, whose critical points are BRST-symmetric. The quenched BRST-symmetric solution was, actually, computed ~\cite{Crisanti04a,Crisanti04e}, and it is {\em unstable} at all free energy levels except the equilibrium one. 
In this Letter we compute the quenched complexity of the marginal, BRST-breaking critical points of the classical TAP free energy, to our knowledge, an problem for forty-five years. We compute it in the two-group formalism below $f^*$, with a 1RSB structure in the space of marginal states, at finite and zero temperature.

The main results are the following.
(i) The quenched complexity is different from the annealed one at $f= f^*$
and extends down to a zero-complexity point $f_0^{\rm 1RSB}$ that, with  respect to $f_0^{\rm ann}$, is
drastically shifted towards the full RSB equilibrium free energy $f_0^{\rm eq}$ 
(cf. Table~\ref{tab:f0}).
(ii) Along the quenched branch, the angle $\gamma$ between the local magnetization and the soft mode of the states grows towards $\pi/2$ as $f \to f_0$, consistent with the scenario in which the lowest  states, the  equilibrium  ones, are still marginal but have soft modes orthogonal to their magnetization.
(iii) The systematic shift of $f_0$ with increasing RSB order,
$f_0^{\rm ann} < f_0^{\rm 1RSB} \lesssim f_0^{\rm eq}$, strongly supports the
conjecture that in the FRSB limit the complexity of marginal states vanishes
exactly at the equilibrium free energy, thereby reconciling the counting of
TAP excited states with Parisi thermodynamics.
We stress that the 2G 1RSB Ansatz is an approximation to the stable 2G FRSB Ansatz, see Appendix~\ref{sec:entropy} for a detailed analysis, therefore its end point at zero complexity does not coincide exactly with the equilibrium ``one group" FRSB solution.

{\em Model and two-group formalism.---}The SK model is defined by the
Hamiltonian $\mathcal{H} = -\sum_{i<j} J_{ij} s_i s_j$, with  $s_i=\pm 1$,
$i\in[1,N]$ and independent Gaussian couplings of zero mean and variance
$\overline{J_{ij}^2}=1/N$. Following Refs.~\cite{Parisi95c,Mueller06} we introduce
the two-group replicated free energy
\begin{equation}
 -\beta m \Phi_{\rm 2G}(m) = \lim_{n\to 0}\lim_{N\to\infty}
 \frac{1}{n N} \ln \overline{Z_J^{mn}},
 \label{eq:Phi2G_def}
\end{equation}
where,  $m$ is the number of ``real'', \`a la Monasson \cite{Monasson95}, replicas per group pair and $n$ the
number of virtual replicas enforcing the quenched average. 
Manipulations in the replica space yield
$\overline{Z_J^{mn}} = \int \mathcal{D}\mathcal{Q}\,
e^{\, n N \mathcal{F}[\mathcal{Q}]}$ with
\begin{equation}
 \mathcal{F}[\mathcal{Q}] = \frac{1}{n}\ln
 {\Tr}_{s}\, e^{\frac{\beta^2}{2}\sum_{ab}^{1,n}\sum_{ij}^{\pm}
 s_a^i \mathcal{Q}^{ij}_{ab} s_b^j}
 + \frac{\beta^2}{4}\Bigl( m - \frac{1}{n}\Tr \mathcal{Q}^2 \Bigr),
 \label{eq:calF}
\end{equation}
where $\mathcal{Q}$ is an $nm\times nm$ overlap matrix. The two-group Ansatz
splits each $m\times m$ block $\mathcal{Q}_{ab}$ into groups of $m+K$ and
$-K$ replicas, with entries parametrized, in the $K\to\infty$ limit, by three
matrices \cite{Mueller06}
\begin{equation}
 \mathcal{Q}^{\pm\mp}_{ab} = Q_{ab}, \quad
 \mathcal{Q}^{\pm\pm}_{ab} = Q_{ab} \pm \frac{A_{ab}}{K} + \frac{C_{ab}}{2K^2}.
 \label{eq:2Gansatz}
\end{equation}
$Q_{ab}$ is the usual overlap between the local magnetizations of states $a$ and
$b$; $A_{ab}$ measures the overlap between the magnetization of state $a$ and
the soft-mode direction of state $b$; $C_{ab}$ is the soft-mode--soft-mode
overlap. 
\begin{figure}[t!]
 \centering
 \includegraphics[width=0.45\columnwidth]{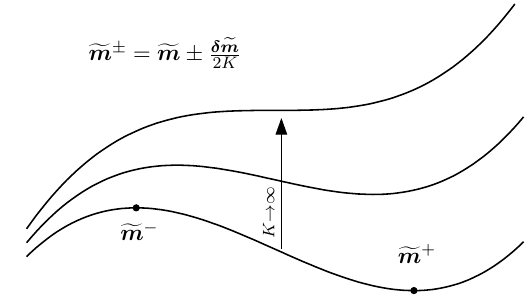}
 \caption{One dimensional pictorial projection in the free energy landscape of the merging into a marginal state  of a conjugated couple of shallow minimum and rank-$1$ saddle. As eventually $K$ is sent to $\infty$ a marginal state is recovered.}
 \label{fig:mergingmarginal}
\end{figure}
A 1D sketch of a couple of minimum with magnetization $\bm{\widetilde{m}}^+$ and rank-$1$ saddle with magnetization $\bm{\widetilde{m}}^-$ in the TAP free energy landscape merging into a marginal state as $K\to\infty$ is plotted in Fig. \ref{fig:mergingmarginal}.

A useful  observable, introduced in Ref.~\cite{Mueller06},
is the angle $\gamma$ between magnetization and soft mode within a state:
\begin{equation}
 \cos\gamma = \frac{\langle \delta \widetilde{m}\, \widetilde{m}\rangle}
 {\sqrt{\langle \delta\widetilde{m}^2\rangle \langle \widetilde{m}^2\rangle}}
 = \frac{A_{aa}}{\sqrt{Q_{aa} C_{aa}}}.
 \label{eq:gamma}
\end{equation}
The complexity is, then, the Legendre transform of the two-group replicated free energy functional
\cite{Monasson95,Bray80,Mueller06}:
\begin{eqnarray}
 \Sigma(f) &=& \max_m \bigl[\beta m f - \beta m \Phi_{\rm 2G}(m)\bigr],
 \label{eq:LT}
 \\
 f(m) &=& \frac{\partial\, [m \Phi_{\rm 2G}]}{\partial m}, 
 \qquad
 \Sigma(m) = m^2 \frac{\partial\, [\beta \Phi_{\rm 2G}]}{\partial m}.
 \label{eq:fm_sigmam}
\end{eqnarray}

{\em Annealed approximation and its instability.---}Retaining only the
diagonal blocks, $Q_{ab}=Q\,\delta_{ab}$, $A_{ab}=A\,\delta_{ab}$,
$C_{ab}=C\,\delta_{ab}$, reproduces through Eqs.~(\ref{eq:LT})--(\ref{eq:fm_sigmam}) the annealed two-group free energy and the Bray--Moore annealed
complexity \cite{Bray80,Crisanti03c,Mueller06}; the explicit saddle-point
equations for $(Q,A,C)$ are recalled in Appendix~\ref{app:EM}.
The annealed solution is exact only where it is locally stable. The relevant replicon eigenvalue of the fluctuation Hessian is \cite{Mueller06}
\begin{equation}
 \Lambda_{R} = 1 - \left[\beta\left(1-Q +A+\frac{mQ}{2}\right)
 + \sqrt{R}\,\right]^2 ,
 \label{eq:replicon}
\end{equation}
with $R \equiv Q\,[\beta^2(C-2A) + \beta^2(m A + Q\, m^2/4)]$. It vanishes at $f^{*}(T)$, which is the Bray--Moore stability condition \cite{Bray81} in the 2G formalism. For $f<f^{*}$, $\Lambda_R<0$, so the annealed approximation fails where its zero-complexity point lies, below the Parisi ground-state energy $e_0^{\rm eq}=-0.76317$ \cite{Parisi79,Crisanti02b,SchmidtOppermann08}. Consistency requires a quenched computation with replica symmetry breaking among the marginal states.

\begin{figure*}[t!]
 \centering
 \includegraphics[width=\textwidth]{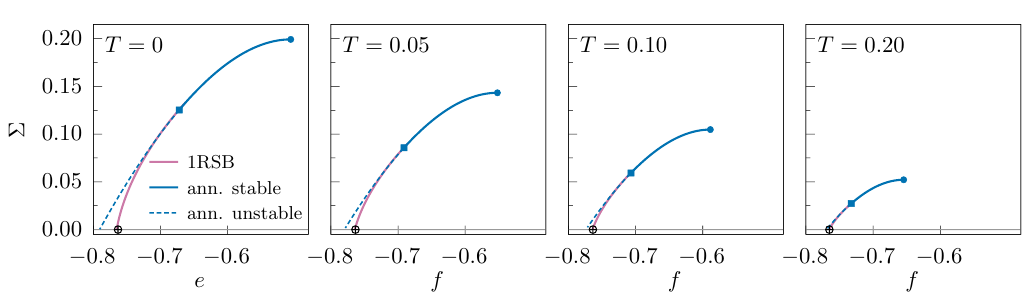}
 \caption{Two-group complexity $\Sigma$ versus free-energy density $f$ in the annealed approximation (blue curve) and quenched 1RSB Ansatz (pink curve), for temperature values $T=0,0.05,0.1,0.2$.
 Full circles mark $f_{\rm max}$ ($m=0$), squares the annealed instability
 point $f^{*}$. Open circles  mark the FRSB equilibrium free-energy   $f_0^{\rm eq}$.  See  Table~\ref{tab:f0} for the main values.}
 \label{fig:Sigma_f}
\end{figure*}

{\em Quenched 1RSB computation.---}We take the matrices $Q_{ab}$, $A_{ab}$,
$C_{ab}$ in the one-step RSB form: diagonal values $(Q,A,C)$, inner-block values
$(Q_1,A_1,C_1)$ within diagonal blocks of size $m_1$, and outer values
$(Q_0,A_0,C_0)$ outside. The saddle-point equations for the external overlaps
only admit the solution $Q_0=A_0=C_0=0$ in zero external field; the RSB structure among marginal states is thus entirely carried
by $(Q_1,A_1,C_1)$ and by the block-size parameter $m_1$.
The resulting quenched 2G 1RSB free energy is
\begin{align}
 &\beta m \Phi_{\rm 2G}^{\rm 1RSB} = -\beta \phi_m^{\rm 1RSB}+\beta^2 A (1-Q)
- \frac{\beta^2m}{4} (1-Q)^2 \nonumber \\
& +\frac{\beta^2}{2} \left[ A^2+QC+2\,m\,QA +\frac{m^2Q^2}{2}\right] 
\label{eq:Phi1RSB}\\
&-\frac{\beta^2 \left( 1-m_1\right)}{2}\left[ A_1^2+Q_1C_1+ 2\,m\,Q_1A_1+\frac{m^2\,Q_1^2}{2}\right], \nonumber
 \end{align}
 with
 \begin{align}
 \beta\phi_m^{\rm 1RSB} &= \frac{1}{m_1} \ln \int
 \mathcal{D}\mu_1
 \left[ \zeta_h(\mu_1;X) \right]^{m_1},
 \label{eq:phim1RSB}
 \\
\zeta_h(\mu_1;X) &\equiv  \int_{-\infty}^{\infty} \frac{dh}{\sqrt{2\pi (Q-Q_1)}}\;
 e^{\mathcal{A}[h,\mu_1;X]} 
 \end{align}
with
 $X\equiv \{m;\,Q,A,C,Q_1,A_1,C_1\}$,$\mu_1\equiv \{z_1,y_1,t_1\}$ and $\mathcal{D}\mu_1= \mathcal{D}z_1 \, \mathcal{D}y_1 \, \mathcal{D}t_1$.
The local action $\mathcal{A}[h,\mu_1; X]$, see Appendix~\ref{app:EM}
[Eq.~(\ref{EM:kernel})], couples the local field $h$ to three Gaussian fields
$(z_1,t_1,y_1)$ through the two combinations
$w_1 =\sqrt{Q_1-A_1}\,z_1 + \sqrt{A_1}\,y_1$, $v_1 =\sqrt{C_1-A_1}\,t_1 + \sqrt{A_1}\,y_1$.
Here, $w_1$ is the frozen
part of the field conjugate to the magnetization and $v_1$ to the soft mode.
Extremizing Eq.~(\ref{eq:Phi1RSB}) yields seven coupled saddle-point
equations for $(Q,Q_1,A,A_1,C,C_1,m_1)$, reported in Appendix~\ref{app:EM}
[Eqs.~(\ref{EM:spQ})--(\ref{EM:spm1})] 
\footnote{Their numerical solution at fixed
$(T,m)$, by iteration over nested Gaussian quadratures, is delicate near the
onset of RSB, where the inner-block equations suffer a loss of
significance. We took care of it using an analytic Gaussian integration-by-parts identity
that recasts the $C_1$ equation into a manifestly well-conditioned form
[Eq.~(\ref{EM:C1stable}) and Appendix~\ref{sec:SPeqs}]}.
The complexity and the free energy of the marginal states then follow
parametrically in $m$ from the 1RSB generalization of
Eq.~(\ref{eq:fm_sigmam}) [Eqs.~(\ref{EM:fm})--(\ref{EM:Sigmam})].

At low temperature to $O(T^2)$, the order parameters have behavior $Q = 1 + w_Q T^2 $,
$A = k_A T + w_A T^2$, $C = k_C T + w_C T^2$, $Q_1 = l_{Q_1} + k_{Q_1} T + w_{Q_1}T^2$,
 $A_1 = k_{A_1}\,T + w_{A_1}T^2$, 
 $C_1 = w_{C_1}\,T^2$.
In the $T\to 0$ limit, with finite $y\equiv\beta m$, the whole scheme turns out to collapse to five closed equations for the overlap parameters $l_{Q_1}$, $k_A$, $k_{A_1}$, $w_{C_1}$,  and $m_1$
[Eqs.~(\ref{EM:T0lq})--(\ref{EM:T0wC1})], allowing a direct
evaluation of the ground-state complexity $\Sigma(e)$ as a function of the energy density, parametrically obtained from Eqs.~(\ref{EM:e(y)})--(\ref{EM:sigma(y)}).

{\em Results.---}
Figure~\ref{fig:Sigma_f} shows the quenched complexity curve $\Sigma(f)$ for four
temperatures with the three characteristic points $f_{\rm max}$
($m=0$, maximum of $\Sigma$), $f^{*}$ (annealed  instability) and
$f_0^{\rm 1RSB}$ ($\Sigma=0$).
The
quenched 1RSB complexity coincides with the annealed one at high free energy,
where the annealed solution is stable and $Q_1$ plays no role. It branches
off continuously in correspondence with the replicon instability $f^{*}$, cf Eq.~\eqref{eq:replicon}.
Below $f^{*}$ the quenched complexity is smaller than the
annealed continuation at the same free energy, and reaches zero at a
higher free energy $f_0^{\rm 1RSB}$.
Table~\ref{tab:f0} collects the zero-complexity points against the
equilibrium free energy $f_0^{\rm eq}$ of
the FRSB Ansatz: at all temperatures, including $T=0$, the
1RSB quenched computation closes most of the unphysical gap
$f_0^{\rm eq}-f_0^{\rm ann}$, leaving a residual mismatch of the same relative order as the residual error of the 1RSB
approximation to the equilibrium free energy itself
\cite{Parisi79,Mezard87}.
\begin{table}[b!]
 \begin{ruledtabular}
 \begin{tabular}{cccc}
 $T$ & $f_0^{\rm ann}$ & $f_0^{\rm 1RSB}$ & $f_0^{\rm eq}$ (FRSB) \\
 \hline
 $0.00$ & $-0.79067$  & $-0.76520$ & $-0.76317$   \\
 $0.05$ & $-0.77979$ & $-0.76429$ & $-0.76333$  \\
 $0.10$ & $-0.77384$ & $-0.76406$ & $-0.76345$   \\
 $0.20$ & $-0.76935$ & $-0.76533$ & $-0.76508$   \\
 \end{tabular}
 \end{ruledtabular}
  \caption{Zero-complexity free energy $f_0$ in the annealed and quenched
 1RSB two-group computations, compared with the equilibrium FRSB free energy
 $f_0^{\rm eq}$ 
 \cite{Crisanti02b,SchmidtOppermann08}. 
 }
 \label{tab:f0}
\end{table}
\begin{figure}[t!]
 \centering
 \includegraphics[width=0.6\columnwidth]{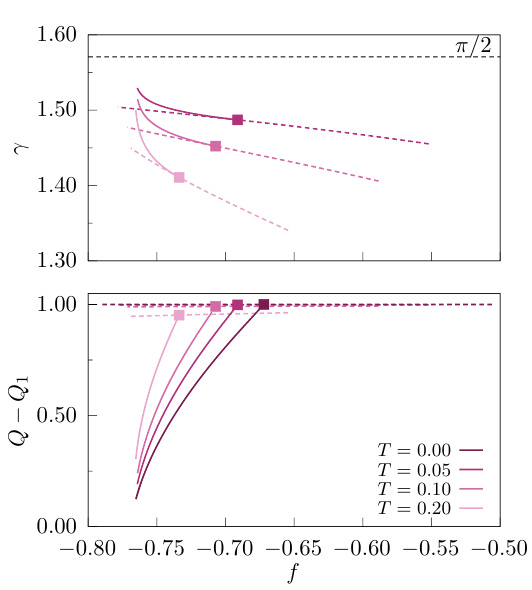}
 \caption{Two-group observables versus the difference in free-energy density between $f$ and its reference zero-complexity value $f_0(T)$, computed in the 1RSB Ansatz and for different values of temperatures. In the first panel the angle between soft modes and magnetizations of marginal states, defined in Eq.~(\ref{eq:gamma}). In the second panel the difference $\Delta Q$ between the diagonal and off-diagonal remaining overlaps.}
 \label{fig:gamma_allT}
\end{figure}

Further observables characterize the lowest marginal states.
First, the overlap difference $Q-Q_1$ decreases markedly along the quenched
branch as $f$ decreases (Fig.~\ref{fig:gamma_allT}): the states that dominate at low free energy
live in increasingly correlated clusters, in an
ultrametric organization of the equilibrium phase.

Second, the soft mode angle
$\gamma$, Eq.~(\ref{eq:gamma}), increases along the quenched branch. This is compatible with  
 $\gamma\to \pi/2$ as $f\to f_0^{\rm eq}$ (Appendix~\ref{app:EMgamma}): this is the scenario in which, in the lowest
marginal states, the soft mode becomes orthogonal to the magnetization and $\Phi_{\rm 2G}\to \Phi_{\rm 1G}\equiv \Phi_{\rm Parisi}$. Indeed, this
is  the property expected for the marginal directions of the exact
equilibrium FRSB states \cite{Parisi79,Mueller06}, for which the soft modes ${\bm{\delta \widetilde{m}}}$
carry no net magnetization component
${\bm {\widetilde{m}}}$. 
The observations that (i) the zero-complexity endpoint
moves towards $f_0^{\rm eq}$   seemingly independent and (ii) the soft mode rotates towards
orthogonality with the state vector, are seemingly independent, and they both substantially improve w.r.t. the annealed computation at any
temperature. 

\begin{figure}
    \centering
    \includegraphics[width=0.6\linewidth]{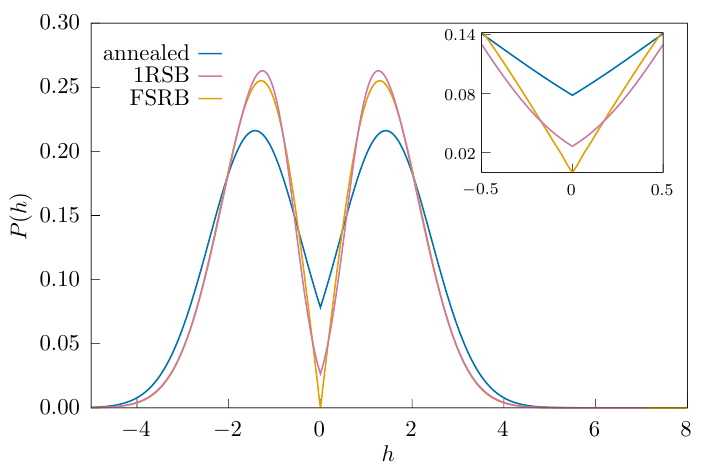}
    \caption{Distribution of local fields $P(h)$ at $T=0$, in the annealed (blue), quenched 1RSB (pink) and full-RSB (yellow) ansatz, evaluated respectively for their zero-complexity point. The distributions are all normalized to unity.}
    \label{fig:fielddist_t0}
\end{figure}

Finally, the local field distribution $P(h)$ of marginal states, which can be obtained as the integrand in $\beta \phi_m^{\rm 1RSB}$, cf. Eq.~(\ref{eq:phim1RSB}), and conveniently marginalized over the Gaussian variables $\mu_1$ and normalized, is shown in Fig.~(\ref{fig:fielddist_t0}). The comparison here is between the $T=0$ local field distribution evaluated in the annealed, in the quenched  1RSB and in the FRSB Ans\"atze at the zero-complexity point.
We pedantically stress that the exact  distribution at the equilibrium is the FRSB one, which indeed exhibits a linear pseudogap at $h=0$ \cite{Anderson1979,Palmer1979,MuellerWyart15}. This analysis shows that the 1RSB Ansatz, quantitatively good, is still an approximation of the exact solution. A more detailed analysis is reported in Appendix~\ref{sec:marginality_notions}.

{\em Discussion.---}The scenario emerging from Table~\ref{tab:f0} and
Figs.~\ref{fig:Sigma_f}--\ref{fig:gamma_allT}--\ref{fig:fielddist_t0} is that the
zero-complexity endpoint of the two-group computation interpolates, as the
order of replica symmetry breaking among marginal states is increased,
between the annealed value $f_0^{\rm ann}$ and the FRSB equilibrium free
energy:
\begin{equation}
 f_0^{\rm ann} \;<\; f_0^{\rm 1RSB} \;<\; f_0^{r{\rm RSB}} \;
 \xrightarrow[r\to\infty]{}\; f_0^{\rm eq}.
 \label{eq:interpolation}
\end{equation}
Looking at the results for the 1RSB approximation 
we can reasonably conjecture that the limit in Eq.~(\ref{eq:interpolation}) is exact: in the
full-RSB two-group theory the complexity of marginal states vanishes exactly
at the equilibrium free energy, i.e., the lowest-lying marginal TAP states
are the equilibrium states of the Parisi solution, and there is no interval
of free energies above $f_0^{\rm eq}$ devoid of states. This closes the
long-standing low energy inconsistency of the annealed Bray--Moore computation
\cite{Bray80,Crisanti03c} and unifies the counting of BRST-breaking TAP states
\cite{Parisi95c,Parisi04b,Mueller06} with equilibrium thermodynamics
\cite{Parisi79}.
%

The shape of
$\Sigma(f)$ at low $f$ has direct consequences for the expected
off-equilibrium dynamics: within a landscape scenario, the free energy
reached at long times after a quench and the aging properties are controlled
by the marginal states that dominate the measure at threshold, and the
$m$ dependence  of the
quenched complexity provides the
weights of those states. 

The tools introduced here
carry over directly to the computation in an external field and to models with mixed 1RSB--FRSB phases \cite{Crisanti04c,Crisanti06,Crisanti07b,TesiAlessandra,TesiTiziana}.
Natural next steps are the computation of the full RSB solution, whose generalized system of antiparabolic Parisi equations was derived already in Refs. \cite{Bray81,Bray1984,Mueller06},
to verify the convergence
pattern of Eq.~(\ref{eq:interpolation}), and the analysis of the fluctuation
spectrum around the quenched two-group saddle point, to identify the marginal
manifold of the state-counting theory itself.
It would also be valuable to compare the quenched two-group solution with the quenched extension of the superspace Kac--Rice formulation announced in Ref.~\cite{KentDobias26}, which provides an independent route to the same BRST-breaking counting, and whose order parameters admit a direct spectral interpretation.
\begin{acknowledgments}
The results leading to this research has received funding from the European Innovation Council (EIC) under the European Union's Horizon Europe research and innovation programme under grant agreement No 101115575 (Q-ONE).
\end{acknowledgments}

\appendix
\makeatletter
\renewcommand{\p@subsection}{\thesection.}
\makeatother

\section{Annealed and quenched 1RSB equations}
\label{app:EM}
{\em Annealed two-group equations.---}%
With diagonal matrices $(Q,A,C)$ the two-group free energy reads
\cite{Mueller06}
\begin{align}
 -\beta m \Phi^{\rm ann}_{\rm 2G} =\;&
 \frac{\beta^2}{4} m (1-Q)^2 - \beta^2 A (1-Q) + \beta\phi_m \nonumber\\
 &- \frac{\beta^2}{2}\Bigl[A^2 + QC + 2 m QA + \frac{m^2 Q^2}{2}\Bigr]
 \label{EM:Phiann}
\end{align}
\begin{align}
 \beta\phi_m =\;& \ln \int_{-\infty}^{+\infty}
 \frac{dh}{\sqrt{2\pi Q}}\,
 \exp\Bigl\{ m \ln [2\cosh\beta h] - \frac{h^2}{2Q} \nonumber\\
 &+ \frac{A}{Q}\beta h \tanh \beta h
 + \frac{\beta^2}{2}\frac{QC-A^2}{Q}\tanh^2\beta h \Bigr\}.
 \label{EM:phiann}
\end{align}
The saddle-point equations, with $\langle\cdot\rangle$ the average over the
normalized measure in Eq.~(\ref{EM:phiann}), are
\begin{align}
 Q &= \langle \tanh^2 (\beta h)\rangle, \nonumber \\
 A &= \frac{1}{2}\left[ -[1+(m-1)Q]
 + \frac{\langle h\tanh(\beta h)\rangle}{\beta Q} \right],
 \label{EM:spann}\\
 C &= 2A + [m(1-Q) + m^2 Q + 2 m A] + \frac{A^2}{Q} \nonumber\\
 &\quad - \frac{2}{\beta Q}\left(m + \frac{A}{Q}\right)
 \langle h \tanh(\beta h)\rangle
 - \frac{1}{\beta^2 Q}\left(1 - \frac{\langle h^2\rangle}{Q}\right).
 \nonumber
\end{align}
The complexity follows from Eq.~(\ref{eq:fm_sigmam}). The replicon,
Eq.~(\ref{eq:replicon}), the Plefka criterion and the spin-glass
susceptibility of the annealed states, together with its $T \to 0$ limit, are analyzed in Appendices~\ref{subsec:stability} and \ref{sec:annT0}.\\
{\em Quenched 1RSB kernel and saddle-point equations.---}%
Defining $\Delta Q = Q-Q_1$, $\Delta A = A-A_1$ and $\Delta C = C-C_1$, the local action  entering in  Eq.~(\ref{eq:Phi1RSB}) is
\begin{align}
 \mathcal{A}[h,\mu_1; X] =\;&
 m \ln[2\cosh\beta h] \nonumber\\
 &+ \frac{\beta^2}{2}\left[\Delta C - \frac{\Delta A^2}{\Delta Q}\right]
 \tanh^2\beta h \nonumber\\
 &- \frac{h^2}{2\Delta Q}
 + \frac{\Delta A}{\Delta Q}\,\beta h \tanh\beta h
 + \frac{w_1 h}{\Delta Q} \nonumber\\
 &+ \beta \left[v_1 - \frac{\Delta A}{\Delta Q} w_1 \right]\tanh\beta h
 - \frac{w_1^2}{2\Delta Q},
 \label{EM:kernel}
\end{align}
where the field $w_1$ and $v_1$ are
\begin{align}
w_1 &=\sqrt{Q_1-A_1}\,z_1 + \sqrt{A_1}\,y_1,
\nonumber
\\
v_1 &=\sqrt{C_1-A_1}\,t_1 + \sqrt{A_1}\,y_1.
\nonumber
\end{align}
Let us define the averages  
\begin{align}
    \avh{X} =\;& \zeta_h^{-1}\!\int\! \frac{dh}{\sqrt{2\pi\Delta Q}} e^{\mathcal{A}} X, \quad \zeta_h = \int\! \frac{dh}{\sqrt{2\pi\Delta Q}} e^{\mathcal{A}}, \nonumber \\
\avmu{Y} =& \zeta_{\mu_1}^{-1}\!\int\!\mathcal{D}\mu_1\,
\zeta_h^{m_1}\, Y, \quad \zeta_{\mu_1}=\int\!\mathcal{D}\mu_1\,
\zeta_h^{m_1}, \nonumber  \\
 \mathcal{D}\mu_1 =& \mathcal{D}z_1\,\mathcal{D}t_1\, \mathcal{D}y_1,
 \label{EM:averages}
\end{align}
and introduce the abbreviations
\begin{align}
 &G = \avmu{\avh{h\tanh\beta h}}, \quad
 G_1 = \avmu{\avh{h}\avh{\tanh\beta h}}, 
 \nonumber\\
 &K = \avmu{\avh{h^2}}, \quad
 K_1 = \avmu{\avh{h}^2}, 
  \label{EM:integrals}
\\
& \chi_X = X - (1-m_1) X_1 \quad X \in \{Q, G, K\},
 \\
 &W_t = \avmu{w_1 \avh{\tanh \beta h}}, 
 W_h = \avmu{w_1 \avh{h}}.
\end{align}
The seven saddle-point equations read
\begin{align}
 Q &= \avmu{\avh{\tanh^2\beta h}},
 \label{EM:spQ}\\
 Q_1 &= \avmu{\avh{\tanh\beta h}^2},
 \label{EM:spQ1}\\
 A &= -\frac{1-Q+m\,Q}{2}
 +\frac{\chi_Q G - \chi_G Q_1}{2\beta \Delta Q\chi_Q}
 \label{EM:spA}\\
 A_1 &= -\frac{Q_1 m}{2}
 + \frac{Q G_1 - Q_1 G}{2\beta \Delta Q\, \chi_Q},
 \label{EM:spA1}\\
 C &= 2A - m(1-Q) - 2 mA - m^2 Q + \frac{I_1}{\beta^2},
 \label{EM:spC}
\end{align}
\begin{align}
 C_1 =& -2 m A_1 - m^2 Q_1 + \frac{\mathcal{N}_1}{\beta^2\,\Delta Q^2},
 \label{EM:C1stable}\\
 \mathcal{N}_1 =& \, K_1 - 2 W_h + \avmu{w_1^2} - 2\beta\Delta A\,(G_1 - W_t) \nonumber \\
&+\beta^2 \Delta A^2 Q_1.
\end{align}
\begin{eqnarray}
 &&\avmu{\ln \zeta_h} - \beta\phi_m^{\rm 1RSB} =\beta^2 m_1 \psi\left(Q_1,A_1,C_1,m\right), \\
 \label{EM:spm1}
 &&\psi(x,y,z,m)\equiv\frac{1}{2}\Bigl[ y^2 + xz + 2 m xy + \frac{m^2 x^2}{2}\Bigr].
 \label{EM:psi}
\end{eqnarray}
The  $I_1$ is an explicit combination of the integrals in
Eq.~(\ref{EM:integrals}), reported in Eq.~(\ref{eq:I1}). 
Eq.~(\ref{EM:C1stable}) is a simplified form of the  saddle point equation for $C_1$ (see Appendix~\ref{sec:SPeqs} for the derivation),  obtained by
an exact Gaussian
integration-by-parts on the outer field $z_1$.
Finally, Eq. (\ref{EM:psi}) is the 
 stationarity condition of $\Phi_{\rm 2G}$, cf. Eq. \eqref{eq:Phi1RSB}, with respect to the RSB breaking parameter $m_1$.  
 
For $Q_1=A_1=C_1=0$ all equations collapse onto the annealed set,
Eq.~(\ref{EM:spann}).

{\em Quenched BRST-broken Complexity.---}In the 2G scheme of computation with one step of RSB, the free energy of the states and the complexity, obtained through Eq.~\ref{eq:fm_sigmam}, are
\begin{align}
 \label{EM:fm}
 f^{\rm 1RSB}(m) =\;& -\frac{\avmu{\avh{\ln 2\cosh\beta h}}}{\beta}
 \\
 &- \frac{\beta}{4}\bigl[(1-Q)^2 - 2 m Q^2 - 4 A Q\bigr] \nonumber\\
 &- \frac{\beta}{4}(1-m_1)\bigl[2 m Q_1^2 + 4 A_1 Q_1\bigr],
 \nonumber
\\
 \Sigma^{\rm 1RSB}(m) =\;& \beta\phi_m^{\rm 1RSB}
 - m \avmu{\avh{\ln 2\cosh\beta h}} 
  \label{EM:Sigmam}
  \\
 &+ \frac{\beta^2}{4}\bigl[m^2 Q^2 - 4 A (1-Q) - 2(A^2+QC)\bigr] \nonumber\\
 &- \frac{\beta^2 (1-m_1)}{4}\bigl[m^2 Q_1^2 - 2(A_1^2 + Q_1 C_1)\bigr].
\nonumber
\end{align}

{\em $T\to 0$ limit.---}
As $T\to 0$ with $y = \beta m$ fixed, the order parameters behave as
$Q = 1 + w_Q T^2$, $A = k_A T + w_A\, T^2+O(T^3)$, $C = k_C T +w_C\,T^2 + O(T^3)$,
$Q_1 = l_{Q_1} + k_{Q_1}\,T+O(T^2)$, $A_1=k_{A_1}\,T+w_{A_1}\,T^2 +O(T^3)$ and $C_1=k_{C_1}\,T+w_{C_1}\,T^2 +O(T^3)$.  Only $k_A$, $l_{Q_1}$, $k_{A_1}$ and $w_{C_1}$ survive in the ground-state theory and their saddle-point equations read
\begin{align}
 l_{Q_1} =\;& \avmu{\avh{\operatorname{sign} h}^2},
 \label{EM:T0lq}\\
 k_A =\;& \frac{1}{1-l_{Q_1}}\Bigl[\avmu{\avh{|h|}}
 - \sqrt{l_{Q_1}}\,\avmu{z_1\avh{\sign h}} \nonumber \\
 &- (k_A-k_{A_1})\Bigr]-y,
 \label{EM:T0kA}\\
 k_{A_1} =\;&  \frac{1}{1-l_{Q_1}}
 \Bigl[\avmu{\avh{h}\avh{\sign h}}
 - \sqrt{l_{Q_1}}\,\avmu{z_1\avh{\sign h}} \nonumber \\
 &- (k_A-k_{A_1})\avmu{\avh{\sign h}^2}\Bigr]-y\,l_{Q_1},
 \label{EM:T0kA1}\\
 w_{C_1} =\;&  \frac{1}{(1-l_{Q_1})^2}
 \Bigl[\avmu{\avh{h}^2}
 - 2\sqrt{l_{Q_1}}\,\avmu{z_1\avh{h}}
 \nonumber\\
 &+ l_{Q_1}\avmu{z_1^2} - 2(k_A-k_{A_1})\avmu{\avh{h}\avh{\sign h}}
 \nonumber\\
 &+ 2(k_A-k_{A_1})\sqrt{l_{Q_1}}\,\avmu{z_1\avh{\sign h}}
 \nonumber\\
 &+ (k_A-k_{A_1})^2\avmu{\avh{\sign h}^2}\Bigr]-2y\,k_{A_1} - y^2 l_{Q_1},
 \label{EM:T0wC1}
\end{align}
while the saddle-point equation for $m_1$ reads
\begin{figure}[t!]
  \centering \includegraphics[width=0.8\columnwidth]{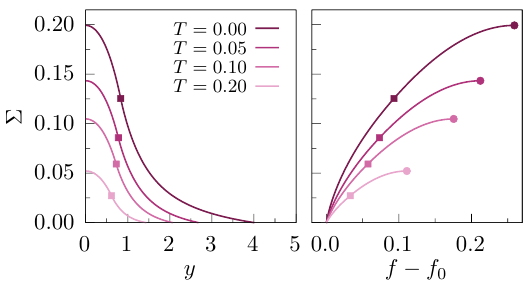}
 \caption{2G quenched 1RSB complexity $\Sigma$ for different values of temperature.  In the left panel, $\Sigma$ versus the rescaled parameter $y=\beta m$. In the right panel, $\Sigma$ versus the difference in free-energy density between $f$ and its reference zero-complexity value $f_0(T)$. Full circles mark $f_{\rm max}$ ($y=0$), squares the annealed stability point $f^{*}$ where the Eq.~(\ref{eq:replicon}) vanishes. }
 \label{fig:sigma_y_allT}
\end{figure}
\begin{align}
\avmu{\ln\zeta_h^0} - (\beta\phi_m^{\rm 1RSB})^0 =\;m_1 \psi(l_{Q_1},k_{A_1},w_{C_1},y),
 \label{eM:T0m1}
\end{align}
where the function $\psi$ is defined in Eq.~(\ref{EM:psi}). In the previous saddle point equations, the definitions of the integrals over $h$ and the  Gaussian variables $z_1$ and $t_1$ are
\begin{align}
 \avh{X} =\;& \frac{1}{\zeta_h^0}\int\frac{dh}{\sqrt{2\pi\,(1-l_{Q_1})}}\,
 e^{\mathcal{A}^0}\,X , \\
     \zeta_h^0 =\;& \int\! \frac{dh}{\sqrt{2\pi(1-l_{Q_1})}} \, e^{\mathcal{A}^0},  \\
\avmu{Y} =\;& \frac{1}{\zeta_{\mu}}\int\mathcal{D}z_1\,\mathcal{D}t_1\,
 [\zeta_h^0]^{m_1}\,Y, \\
 \zeta_{\mu}=\;&\int\mathcal{D}z_1\,\mathcal{D}t_1\,[\zeta_h^0]^{m_1}.
\end{align}
The explicit form of the replicated free-energy functional and $\mathcal{A}^0(h,z_1, t_1;\, m,k_A,l_{Q_1}, k_{A_1},w_{C_1})$ are given in Appendix~\ref{sec:T0}.

The energy of the states and the ground-state complexity read
\begin{align}
 e(y) =\;& k_A - \avmu{\avh{|h|}} 
 \nonumber
 \\
 & + \frac y2
 - (1-m_1)\left(l_{Q_1}k_{A_1} + \frac{y\,l_{Q_1}^2}{2}\right),
 \label{EM:e(y)}\\
 \Sigma(y)=\;& -\frac{k_A^2}{2} - \frac{(k_A-k_{A_1})^2}{2\,(1-l_{Q_1})} - \frac{w_{C_1}}{2}
 - y\,\avmu{\avh{|h|}} \nonumber \\
 &+ \frac{y^2}{4}+ \frac{1-m_1}{2}\Bigl[k_{A_1}^2 + l_{Q_1}w_{C_1}
 - \frac{y^2 l_{Q_1}^2}{2}\Bigr] + \frac{\ln \zeta_{\mu}}{m_1}.
 \label{EM:sigma(y)}
\end{align}
Setting $l_{Q_1}=k_{A_1}=w_{C_1}=0$ recovers the annealed theory of
Bray and Moore at $T=0$ \cite{Bray80,Mueller06} . Figure~(\ref{fig:sigma_y_allT}) shows the quenched complexity as a function of
$y=\beta m$ (left) and of the free-energy difference $f-f_0$ (right), where $f_0$ is the zero-complexity, at
various temperatures including $T=0$ where $f-f_0=e-e_0$.


\section{The two-group replicated free energy}
\label{sec:2G}

The SK Hamiltonian is $\mathcal{H} = -\sum_{i<j} J_{ij} s_i s_j$ with Gaussian
couplings of variance $1/N$. The two-group replicated free energy is defined
as
\begin{equation}
 -\beta m \Phi_{\rm 2G}(m) = \lim_{n\to0}\lim_{N\to\infty}
 \frac{1}{nN}\ln \overline{Z_J^{mn}}.
\end{equation}
Averaging over the disorder,
\begin{equation}
 \overline{Z_J^{mn}} = \sum_{\{s\}} \exp\left\{ \frac{\beta^2}{4N}
 \left[ m n N^2 + 2 \sum_{\alpha<\beta}\Bigl(\sum_i s_i^\alpha s_i^\beta
 \Bigr)^2 \right]\right\},
\end{equation}
and after $nm(nm-1)/2$ Hubbard--Stratonovich transformations one obtains
$\overline{Z_J^{mn}} = \int \mathcal{D}\mathcal{Q}\,
e^{nN\mathcal{F}[\mathcal{Q}]}$ with
\begin{equation}
 \mathcal{F}[\mathcal{Q}] = \frac{1}{n} \ln \sum_{\{s_a^i\}}
 \exp\Bigl( \frac{\beta^2}{2}\sum_{ab}\sum_{ij} s_a^i \mathcal{Q}^{ij}_{ab}
 s_b^j\Bigr) + \frac{\beta^2}{4}\Bigl(m - \frac1n \Tr \mathcal{Q}^2\Bigr),
 \label{eqSM:calF}
\end{equation}
where $\mathcal{Q}$ is an $nm\times nm$ matrix with block indices
$a,b\in[1,n]$ and inner indices $i,j\in[1,m]$. The two-group Ansatz
\cite{Bray78,Mueller06} divides each $m\times m$ block $\mathcal{Q}^{ij}_{ab}$
into sub-blocks of sizes $(m+K)$ and $(-K)$,
\begin{equation}
 \mathbf{Q}_{ab} = \begin{pmatrix}
 Q^{++}_{ab} & Q^{+-}_{ab}\\[2pt] Q^{-+}_{ab} & Q^{--}_{ab}
 \end{pmatrix}, \qquad
 Q^{+-}_{ab} = Q_{ab}, \qquad
 Q^{\pm\pm}_{ab} = Q_{ab} \pm \frac{A_{ab}}{K} + \frac{C_{ab}}{2K^2},
 \label{eqSM:matrices}
\end{equation}
in the limit $K\to\infty$. The matrix $Q_{ab}$ is the overlap between the
magnetizations of states $a$ and $b$; $A_{ab}$ the overlap between the
magnetization of state $a$ and the soft-mode direction of state $b$; $C_{ab}$
the soft-mode--soft-mode overlap \cite{Mueller06}. Following
Ref.~\cite{Mueller06}, the angle between the magnetization
vector of a state and the soft mode of the state labeled $b$ is defined by
\begin{equation}
 \cos(\gamma_{ab}) = \frac{\langle \delta \widetilde{m}^a_i\, \widetilde{m}^b_i\rangle}
 {\sqrt{\langle \delta \widetilde{m}_i^2\rangle\langle \widetilde{m}_i^2\rangle}}
 = \frac{A_{ab}}{\sqrt{Q_{aa}C_{aa}}},
 \label{eq:gammadef}
\end{equation}
the soft modes being normalized by $\delta \widehat{\widetilde{m}}_i =
\delta m_i/\sqrt{\langle \delta \widetilde m_i^2\rangle} = \delta \widetilde m_i/\sqrt{C_{aa}}$,
since the two-group construction fixes the {\em direction} but not the
magnitude of $\delta \widetilde  m_i$. Throughout we quote the diagonal value
$\gamma \equiv \gamma_{aa}$.

In terms of these matrices  in Eq. \eqref{eqSM:matrices}, the two-group free energy takes the form
\begin{equation}
 -\beta m \Phi_{\rm 2G} = \lim_{n\to 0}\left[
 \beta\phi_m + \frac{\beta^2}{4} m (1-Q)^2 - \beta^2 A (1-Q)
 - \frac{\beta^2}{n}\sum_{ab}\Bigl( \frac12\bigl(A_{ab}^2 + Q_{ab}C_{ab}\bigr)
 + m Q_{ab} A_{ab} + \frac{m^2}{4} Q_{ab}^2 \Bigr)\right],
 \label{eq:Phi_first}
\end{equation}
with the single-site term
\begin{align}
 e^{n\beta\phi_m} =\;& 2^{nm} \int_{-i\infty}^{+i\infty} \prod_a
 \frac{dx_a}{2\pi i} \int_{-1}^{1}\prod_a \frac{d\widetilde m_a}{(1-\widetilde{m}_a^2)}
 \exp\Bigl\{ -\sum_a \Bigl[ x_a \tanh^{-1} \widetilde m_a + \frac{m}{2}\ln(1-\widetilde m_a^2)
 \Bigr] \nonumber\\
 &+ \beta^2 \sum_{ab}\Bigl[ \frac12 x_a Q_{ab} x_b
 + \frac12 \widetilde m_a C_{ab} \widetilde m_b + \widetilde m_a A_{ab} x_b \Bigr]\Bigr\}.
 \label{eq:expphi}
\end{align}
The variables $\widetilde m_a$ and $x_a$ are the local magnetization of state $a$ and
its conjugate variable, whose correlations with $\widetilde m_b$ define the soft-mode
overlaps; the direct connection with the counting of solutions of the TAP
equations \cite{Thouless77} and with the Bray--Moore action \cite{Bray80,Bray81} is
established in Ref.~\cite{Mueller06} and recalled in Appendix~\ref{subsec:stability}.
The complexity is obtained by the Legendre transform \cite{Monasson95,Mueller06}
\begin{equation}
 \Sigma_{LT}(f) = \max_m \bigl[\beta m f - \beta m \Phi_{\rm 2G}(m)\bigr],
 \qquad
 f(m,\beta) = \frac{\partial (m\Phi_{\rm 2G})}{\partial m}, \qquad
 \Sigma(m,\beta) = m^2 \frac{\partial (\beta \Phi_{\rm 2G})}{\partial m}.
 \label{eqSM:LT}
\end{equation}

\section{Annealed approximation}
\label{sec:annealed}

\subsection{Saddle-point equations and complexity}
\label{subsec:spe_annealed}
In the annealed approximation only the diagonal entries
of $Q, A, C$ are retained; in the two-group framework, this corresponds to give the matrices diagonal structure $X_{ab}=X\delta_{ab}$, for $X \in \{Q,A,C\}$. Performing the $x_a$ Gaussian integration and with the change of variable
$\widetilde m_a = \tanh\beta h_a$, the two-group replicated free-energy reads
\begin{equation}
 -\beta m \Phi^{\rm ann}_{\rm 2G} =
 \frac{\beta^2}{4} m(1-Q)^2 - \beta^2 A (1-Q)
 - \frac{\beta^2}{2}\Bigl[ A^2 + QC + 2 m QA + \frac{m^2 Q^2}{2}\Bigr]
 + \beta\phi_m^{\rm ann},
\end{equation}
\begin{equation}
 \beta\phi_m^{\rm ann} = \ln \int_{-\infty}^{+\infty}
 \frac{dh}{\sqrt{2\pi Q}} \exp\left\{ m \ln(2\cosh\beta h)
 - \frac{h^2}{2Q} + \frac{A}{Q}\beta h \tanh\beta h
 + \frac{\beta^2}{2}\,\frac{QC - A^2}{Q}\, \tanh^2\beta h \right\}.
 \label{eq:Pann}
\end{equation}
Denoting by $\langle\cdot\rangle$ averages over the normalized measure
$\mathcal{P}^{\rm ann}(h)$ defined by the integrand of Eq.~(\ref{eq:Pann}),
the stationarity conditions in $C$, $A$ and $Q$ give
\begin{align}
 Q &= \langle \tanh^2(\beta h)\rangle, \\
 A &= \frac12\Bigl[ -[1+(m-1)Q] + \frac{1}{\beta Q}\langle h \tanh(\beta h)
 \rangle \Bigr],\\
 C &= 2A + \bigl[ m(1-Q) + m^2 Q + 2 m A \bigr] + \frac{A^2}{Q}
 - \frac{2}{\beta Q}\Bigl( m + \frac AQ\Bigr)\langle h\tanh(\beta h)\rangle
 - \frac{1}{\beta^2 Q}\Bigl( 1 - \frac1Q \langle h^2\rangle \Bigr),
\end{align}
solved iteratively at fixed $(T,m)$. The parametric free energy and
complexity, Eq.~(\ref{eqSM:LT}), read
\begin{align}
 f^{\rm ann}(m) &= -\frac{\langle \ln 2\cosh\beta h\rangle}{\beta}
 - \frac{\beta}{4}\Bigl[ (1-Q)^2 - 2 m Q^2 - 4 A Q\Bigr],\\
 \Sigma^{\rm ann}(m) &= \beta\phi_m^{\rm ann}
 - m \langle \ln 2 \cosh\beta h\rangle
 + \frac{\beta^2}{4}\bigl[ m^2 Q^2 - 4 A (1-Q) - 2 (A^2 + QC)\bigr].
\end{align}
Setting $A=C=0$ recovers the unstable BRST--symmetric (``one-group") solution, whose
free energy coincides with the standard 1RSB equilibrium free energy of the
SK model with $q_0=0$, $q_1 = Q$ and breaking parameter $m$
\cite{Cavagna03b,Crisanti03c}; its complexity vanishes identically at $m=0$.

\subsection{Stability: replicon, Plefka criterion, spin-glass susceptibility}
\label{subsec:stability}

The two-group annealed solution, that is the Bray--Moore complexity \cite{Bray80}, is exact for $f^* < f < f_{\rm max}$, where
$f_{\rm max}$ corresponds to $m=0$ (maximum of $\Sigma$) and $f^*$ marks the
vanishing of the replicon eigenvalue of the Hessian of the fluctuations
around the annealed saddle point. In two-group variables
\cite{Bray81,Mueller06}, it is
\begin{equation}
 \Lambda_R = 1 - \left[ \beta(1-Q) + \beta\Bigl(A + \frac{mQ}{2}\Bigr)
 + \sqrt{Q\Bigl[ \beta^2 (C-2A) + \beta^2\Bigl(mA + \frac{m^2}{4}\Bigr)
 \Bigr]}\,\right]^2,
 \label{eqSM:replicon}
\end{equation}
whose vanishing,
\begin{equation}
 1 = \beta(1-Q) + \beta\Bigl(A + \frac{mQ}{2}\Bigr)
 + \sqrt{Q\Bigl[\beta^2(C-2A) + \beta^2\Bigl(mA + \frac{m^2}{4}\Bigr)\Bigr]},
 \label{eq:criticalcondition}
\end{equation}
is the exact counterpart of Eq.~(A5.14) of Bray and Moore \cite{Bray81}. At
high $f$ the argument of the square root is negative (complex pair of
eigenvalues, no instability); then $\Lambda_R$ becomes real positive and vanishes at $f = f^*$. Below that free energy threshold the annealed solution is unstable. The internal consistency of the states can be further monitored via the
Plefka criterion \cite{Plefka82,Plefka02} for the TAP Hessian
$\chi^{-1}_{ij} = \partial^2 F_{\rm TAP}/\partial m_i \partial m_j$,
\begin{equation}
 x_P \equiv 1 - \beta^2 \frac1N \sum_i (1-m_i^2)^2 \geq 0
 \quad\longrightarrow\quad
 x_P = 1 - \beta^2\bigl( 1 - 2\langle \tanh^2\beta h\rangle
 + \langle \tanh^4\beta h\rangle\bigr),
\end{equation}
and the associated spin-glass susceptibility
\begin{equation}
 \chi_{SG} \equiv \frac1N \sum_{ij}\chi^2_{ij} = \frac{1-x_P}{x_P},
\end{equation}
which diverges as $x_P\to 0$. Within the annealed approximation $x_P>0$ on
the whole physical branch and $\chi_{SG}$ grows as $f$ decreases; at $T=0$, $x_P\to1$ and
$\chi_{SG}\to 0$.
\subsection{$T\to 0$ limit of the annealed theory}
\label{sec:annT0}
As $T \to 0$ the saddle-point values behave as $Q = 1 + w_Q T^2 + O(T^3)$,
$A = k_A T + w_A T^2+O(T^3)$, $C = k_C T + w_C T^2+O(T^3)$. With $y \equiv m\beta$ fixed,
$\lim_{\beta\to\infty}\beta m \Phi^{\rm ann}_{\rm 2G} = \psi^0_{2G}(y)$ depends
on the sole parameter $k_A$,
\begin{eqnarray}
 -\psi^0_{2G}(y)
   &=& \ln \int_{-\infty}^{+\infty}\frac{dh}{\sqrt{2\pi}}\,
   \exp\Bigl\{-\frac{h^2}{2} + (y+k_A)\,|h|\Bigr\}
   - k_A^2 - y\,k_A - \frac{y^2}{4},
   \nonumber\\
 &=& \ln\Bigl[1 - {\rm Erf}\Bigl(-\frac{y+k_A}{\sqrt2}\Bigr)\Bigr]
   + \frac{y^2}{4} - \frac{k_A^2}{2},
   \label{SM:bigPhi0}
\end{eqnarray}
and the unique saddle-point equation reads
\begin{equation}
 k_A = \frac{\langle |h|\rangle - y}{2},
 \label{eq:SPkA}
\end{equation}
the average being taken over the normalized measure $\mathcal{P}^{\rm ann}_0(h)$
defined by the integrand of Eq.~(\ref{SM:bigPhi0}).

Both the measure and Eq.~(\ref{SM:bigPhi0}) depend on $y$ and $k_A$ through the
single combination $u \equiv y + k_A$. It is therefore convenient to introduce
\begin{equation}
 u \equiv y + k_A,
 \qquad
 M(u) \equiv \sqrt{\frac{2}{\pi}}\,
 \frac{e^{-u^{2}/2}}{{\rm Erfc}\bigl(-u/\sqrt{2}\bigr)},
 \label{eq:udefM}
\end{equation}
where ${\rm Erfc}(-u/\sqrt2) = 1 - {\rm Erf}(-u/\sqrt2)$ is the same
combination already appearing in Eq.~(\ref{SM:bigPhi0}). In terms of $u$ the field distribution
and its normalization read
\begin{equation}
 \mathcal{P}^{\rm ann}_0(h) = \frac{1}{\mathcal{N}_0}\,e^{-h^2/2 + u\,|h|},
 \qquad
 \mathcal{N}_0 = \sqrt{2\pi}\;e^{u^{2}/2}\,{\rm Erfc}\bigl(-u/\sqrt{2}\bigr),
 \label{eq:P0ann}
\end{equation}
so that all averages are Gaussian moments truncated to $h>0$; in particular
\begin{equation}
 \langle |h| \rangle = u + M(u),
 \qquad
 {\rm Var}(h) = 1 - u\,M(u) - M(u)^{2}.
 \label{eq:momentsT0}
\end{equation}

Substituting Eq.~(\ref{eq:SPkA}) into the definition of $u$ gives
$u = (y + \langle|h|\rangle)/2$, that is $\langle|h|\rangle = 2u - y$;
comparing with Eq.~(\ref{eq:momentsT0}), the whole annealed $T=0$ theory
collapses onto a single scalar fixed point,
\begin{equation}
 u = y + M(u),
 \qquad\text{equivalently}\qquad
 k_A = M(u),
 \qquad
 \langle|h|\rangle = y + 2k_A .
 \label{eq:SPu}
\end{equation}
The conjugate parameter is thus the $M(u)$ evaluated at the saddle
point.

Once $u$ is known, the replicated free energy,
the energy of the states and the complexity are explicit functions of $u$,
\begin{align}
 -\psi^0_{2G}(y) &= \ln {\rm Erfc}\Bigl(-\frac{u}{\sqrt2}\Bigr)
 + \frac{y^{2}}{4} - \frac{(u-y)^{2}}{2},
 \label{eq:psi0u}\\
 e(y) &= \frac{y}{2} - u,
 \label{eq:eu}\\
 \Sigma(y) &= \ln {\rm Erfc}\Bigl(-\frac{u}{\sqrt2}\Bigr)
 - \frac{u^{2}}{2} + \frac{y^{2}}{4},
 \label{eq:Sigmau}
\end{align}
the terms in $k_A^2$ and $y\,k_A$ cancelling identically against those coming
from $\langle|h|\rangle$.
The complexity is maximal at $y=0$, where
$u = k_A = 0.5060545$ and $\Sigma_{\rm max} = 0.1992285$, and vanishes at
$y_0 = 1.2679549$, where $e_0 = -0.7906684$.
Throughout, averages are taken over the full real line $h\in(-\infty,+\infty)$,
whereas Bray and Moore \cite{Bray81} use the half-line $h\in[0,+\infty)$, on
which $\langle h\rangle = \langle|h|\rangle$ and ${\rm Var}(h)$ is given by
Eq.~(\ref{eq:momentsT0}). In that representation the marginality condition is
\begin{equation}
 1 = \frac{\langle h\rangle}{2}
 + \sqrt{{\rm Var}(h) + \frac{\langle h\rangle^{2}}{4} - 1},
 \label{eq:margBM}
\end{equation}
coinciding with Eq.~(A5.16) of Ref.~\cite{Bray81}. On the saddle point the
radicand collapses to a difference of squares,
${\rm Var}(h) + \langle h\rangle^{2}/4 - 1 = y^{2}/4 - k_A^{2}$, so that
Eq.~(\ref{eq:margBM}) becomes
\begin{equation}
 1 = u - \frac{y}{2} + \sqrt{\frac{y^{2}}{4} - k_A^{2}} .
 \label{eq:margu}
\end{equation}
The square root is real only for $k_A \le y/2$, that is $u \le 3y/2$: this is
the $T=0$ counterpart of the statement made below
Eq.~(\ref{eq:criticalcondition}) that at high $f$ the radicand is negative and
no instability occurs. The threshold $k_A = y/2$, where $e = -y$ identically,
is reached at $y = 0.6239128$, and Eq.~(\ref{eq:margu}) is satisfied at
$y^{*} = 0.8320668$, corresponding to $e^{*} = -0.6721330$, in agreement with
the $T=0$ limit $\epsilon^{*} = -0.672$ of the annealed threshold $f^{*}$
quoted in Ref.~\cite{Mueller06}.
\section{Two-group quenched replica-symmetric Ansatz}
\label{sec:qRS}
As a preliminary step towards RSB we consider the quenched replica-symmetric
(RS) structure $X_{ab} = X\,\delta_{ab} + X_0 (1-\delta_{ab})$ for
$X \in \{Q, A, C\}$, as
\begin{equation}
    Q^{\rm RS}=\begin{pmatrix}
        Q & \, & \, & \,&\, & \\
        \, & Q & \, & \, & Q_0&\, \\
        \, & \, & Q & \, &\, \\
        \, & \, & \, & Q&\, &\,\\
        \, & Q_0 & \, & \,& Q&\, \\
        \, & \, & \, & \,&\, & Q\,
    \end{pmatrix}.
\end{equation}
The final RS replicated free-energy is written in terms of three Gaussian field $\mu_0=(z_0,t_0,y_0)$, combined as 
\begin{equation}
    w_0 \equiv \sqrt{Q_0-A_0}\,z_0+\sqrt{A_0}\, y_0,  \qquad  v_0 \equiv \sqrt{C_0-A_0}\,t_0+\sqrt{A_0}\, y_0, 
\end{equation}
and the overlap parameters $X=\{m;Q_0,A_0,C_0\}$, reads
\begin{eqnarray}
-\beta m \Phi_{\rm 2G}^{RS}
&=&  \frac{\beta^2}{4} m(1-Q)^2
-\beta^2 A(1-Q)-\frac{\beta^2}{2} \left[ A^2+QC+2\,m\,QA +\frac{m^2Q^2}{2}\right]+ \notag \\
&+&\frac{\beta^2}{2}\left[ A_0^2+Q_0C_0+ 2\,m\,Q_0A_0+\frac{m^2\,Q_o^2}{2}\right] + \beta \phi_m^{\rm RS}, \\
\beta \phi_m^{\rm RS}&=&\int \mathcal{D}\mu_0\; \ln \int \frac{dh}{\sqrt{2 \pi (Q-Q_0)}}\; e^{\mathcal{A}^{\rm RS}}[h,\mu_0,;X],
\label{SM:bigPhi_RS}
\end{eqnarray}
with
\begin{align}
 \mathcal{A}^{\rm RS}[h,\mu_0; X] =\;&
 m \ln[2\cosh\beta h]  - \frac{h^2}{2\Delta Q}+\frac{\beta^2}{2}\Bigl[(C-C_0) - \frac{(A-A_0)^2}{Q-Q_0}\Bigr]
 \tanh^2\beta h 
 + \frac{A-A_0}{Q-Q_0}\,\beta h \tanh\beta h \nonumber \\
 &+ \frac{w_0 h}{Q-Q_0} + \beta \Bigl[v_0 - \frac{A-A_0}{Q-Q_0} w_0 \Bigr]\tanh\beta h
 - \frac{w_0^2}{2 (Q-Q_0)}.
\end{align}
The saddle-point equations for the external parameters admit
$Q_0 = A_0 = C_0 = 0$ as the solution in zero field: $Q_0 = \int \mathcal{D}\mu_0\; \langle \tanh\beta h\rangle_h^2$, where the average $\avh{\cdot}$ is computed over the normalized measure defined by the integrand in $h$ of Eq. (\ref{SM:bigPhi_RS}), has no nontrivial solution
without external field, so that the quenched RS scheme collapses onto the
annealed one. Any nontrivial quenched structure must arise at the level of
replica symmetry breaking, as computed in the next section.

\section{Two group quenched 1RSB Ansatz}
\label{sec:q1RSB}

\subsection{Free energy functional}
\label{subsec:q1rsb_freeenergy}
We now give the matrices $Q_{ab}, A_{ab}, C_{ab}$ a one-step RSB structure:
diagonal $(Q,A,C)$, inner-block values $(Q_1,A_1,C_1)$ within diagonal
blocks of size $m_1$, and $(Q_0,A_0,C_0)$ outside, as
\begin{equation}
 Q^{\rm 1RSB} = \begin{pmatrix}
 Q & Q_1 & Q_1 & & & \\
 Q_1 & Q & Q_1 & & Q_0 & \\
 Q_1 & Q_1 & Q & & & \\
 & & & Q & Q_1 & Q_1\\
 & Q_0 & & Q_1 & Q & Q_1\\
 & & & Q_1 & Q_1 & Q
 \end{pmatrix}.
\end{equation}
The general functional involves six Gaussian fields
$\mu_0 = (z_0, t_0, y_0)$ and $\mu_1 = (z_1, t_1, y_1)$ through
\begin{eqnarray}
    w_0 &\equiv& \sqrt{Q_0-A_0}\,z_0+\sqrt{A_0}\, y_0,  \qquad  v_0 \equiv \sqrt{C_0-A_0}\,t_0+\sqrt{A_0}\, y_0, \\
    w_1 &\equiv & \sqrt{\left( Q_1 -Q_0 \right) -\left( A_1 - A_0 \right) }\, z_1 + \sqrt{A_1-A_0}\, y_1, \qquad v_1 \equiv  \sqrt{\left( C_1 -C_0 \right) -\left( A_1 - A_0 \right) }\, t_1 + \sqrt{A_1-A_0}\, y_1 .
\end{eqnarray}
The saddle-point equation for the external overlap $Q_0= \int \mathcal{D}\mu_0 \avmu{\avh{\tanh \beta h}}^2$, where the definitions of the averages will be clearified in Appendix~\ref{sec:SPeqs}, only admits $Q_0=0$ solution in zero external field. This enforces $A_0 = C_0 = 0$ in zero field and the functional then
reduces to Eqs.~(\ref{eq:Phi1RSB})--(\ref{eq:phim1RSB}), with kernel [Eq.~(\ref{EM:kernel})], being
\begin{eqnarray}
 \mathcal{A}[h,\mu_1;X] &=& m \ln(2\cosh\beta h)
 + \frac{\beta^2}{2}\Bigl[ \Delta C - \frac{\Delta A^2}{\Delta Q}\Bigr]
 \tanh^2\beta h - \frac{h^2}{2\Delta Q}
 + \frac{\Delta A}{\Delta Q}\beta h \tanh\beta h
\nonumber
\\
&& + \frac{w_1 h}{\Delta Q}
 + \beta\Bigl[ v_1 - \frac{\Delta A}{\Delta Q} w_1\Bigr]\tanh\beta h
 - \frac{w_1^2}{2\Delta Q},
 \label{eq:kernel}
\end{eqnarray}
where $\Delta Q \equiv Q - Q_1$, $\Delta A \equiv A - A_1$ and $\Delta C \equiv C-C_1$.
\subsection{Saddle-point equations}
\label{sec:SPeqs}
Averages are defined as
\begin{eqnarray}
 \avh{X} &=& \frac{1}{\zeta_h}\int \frac{dh}{\sqrt{2\pi\Delta Q}}\,
 e^{\mathcal{A}}\, X, \quad
 \zeta_h = \int \frac{dh}{\sqrt{2\pi\Delta Q}}\, e^{\mathcal{A}}, 
 \\
 \avmu{Y} &=& \frac{1}{\zeta_{\mu_1}}\int \mathcal{D}\mu_1 \left(\zeta_h\right)^{m_1} \, Y, \quad
 \zeta_{\mu_1} = \int \mathcal{D}\mu_1 \left(\zeta_h\right)^{m_1}, \\
 \mathcal{D}\mu_1 &=& \mathcal{D}z_1 \mathcal{D}t_1
 \mathcal{D}y_1,
\end{eqnarray}
and for a generic order parameter $X$,
\begin{equation}
 \frac{\partial (\beta\phi_m^{\rm 1RSB})}{\partial X} =
 \begin{cases}
 \avmu{\avh{ d\mathcal{A}/dX}} & X \in \{A, A_1, C, C_1\}, \\[2pt]
 \avmu{\avh{ d\mathcal{A}/dQ }} - \tfrac{1}{2\Delta Q} & X = Q, \\[2pt]
 \avmu{\avh{ d\mathcal{A}/dQ_1 } }+ \tfrac{1}{2\Delta Q} & X = Q_1 
 \end{cases}
\end{equation}
holds.
Derivatives of the averages with respect to the outer Gaussian fields can be
reduced using the following identity, which is valid for any observable $X$,
\begin{equation}
\left \langle l \avh{X}\right\rangle_{\mu_1} =
 \avmu{\avh{ \tfrac{d\mathcal{A}}{dl}\, X }}
 - (1-m_1) \left\langle \avh{\tfrac{d\mathcal{A}}{dl}}\avh{X}
 \right\rangle_{\mu_1}, \qquad l \in \{z_1, t_1, y_1\},
 \label{eq:IBPgeneric}
\end{equation}
which follows from integration by parts on the reweighted measure
$\mathcal{D}\mu_1 \zeta_h^{m_1}$. Therefore, introducing the basic integrals
\begin{equation}
 L_t = \langle l_1 \avh{\tanh\beta h}\rangle_{\mu_1}, \quad
 L_h = \langle l_1 \avh{h}\rangle_{\mu_1}, \qquad  l \in \{z_1, t_1, y_1\},
 \label{eq:L}
\end{equation}
\begin{equation}
 W_t = \langle w_1 \avh{\tanh\beta h}\rangle_{\mu_1}, \quad
 W_h = \langle w_1 \avh{h}\rangle_{\mu_1}, 
\end{equation}
\begin{equation}
 G = \avmu{\avh{h\tanh\beta h}}, \quad
 G_1 = \avmu{\avh{h}\avh{\tanh\beta h}}, \quad
 K = \avmu{\avh{h^2}}, \quad
 K_1 = \avmu{\avh{h}^2},
\end{equation}
and their combination
\begin{equation}
\chi_X = X - (1-m_1) X_1, \qquad X \in \{Q, G, K\},
\end{equation}

the seven saddle-point equations are:
\begin{align}
 Q &= \avmu{\avh{\tanh^2\beta h}},
 \label{eq:spQ}\\
 Q_1 &= \langle \avh{\tanh\beta h}^2\rangle_{\mu_1},
 \label{eq:spQ1}\\
 A &= \frac12\left[ -[1-(1-m)Q]
 + \frac{1}{\beta\Delta Q}\Bigl( G - \frac{\chi_G Q_1}{\chi_Q}\Bigr)\right],
 \label{eq:spA}\\
 A_1 &= -\frac{Q_1 m}{2}
 + \frac{1}{2\beta \Delta Q\, \chi_Q}\,\bigl( Q G_1 - Q_1 G \bigr),
 \label{eq:spA1}\\
 C &= 2A - m(1-Q) - 2m A - m^2 Q + \frac{I_1}{\beta^2},
 \label{eq:spC}\\
 C_1 &= -2 m A_1 - m^2 Q_1 + \frac{I_1 + I_2}{\beta^2 (1-m_1)}
 \qquad \text{(raw form)},
 \label{eq:spC1raw}\\
 \avmu{\ln \zeta_h} - \beta\phi_m^{\rm 1RSB} &=
 \frac{\beta^2 m_1}{2}\left[ A_1^2 + Q_1 C_1 + 2 m Q_1 A_1
 + \frac{m^2 Q_1^2}{2}\right],
 \label{eq:spm1}
\end{align}
where
\begin{align}
 I_1 &= \frac{K - 2 W_h + \avmu{w_1^2}
 - 2\beta\Delta A\, G + 2\beta\Delta A\, W_t
 + \beta^2 \Delta A^2 Q}{\Delta Q^2} - \frac{1}{\Delta Q},
 \label{eq:I1}\\
 I_2 &= \frac{\avmu{z_1 w_1} - Z_h + \beta \Delta A\, Z_t}
 {\Delta Q\,\sqrt{Q_1 - A_1}}.
 \label{eq:I2}
\end{align}

The integral $I_1$ depends only on $W_h$, $W_t$ and $\avmu{w_1^2}$, while $I_2$ depends on $Z_h$, $Z_t$ and $\avmu{z_1\,w_1}$ which can be wisely re-arranged as combinations of  $W_h$, $W_t$ and $\avmu{w_1^2}$. This recast is motivated by the numerical solution of the saddle point equations, which is carried out in terms of the variables $w_1$ and $v_1$, whose details are in Appendix~\ref{sec:numerics}. The first two quantities $Z_h$ and $Z_t$ are evaluated directly from Eq. (\ref{eq:L}) using 
\begin{equation}
    \partial_{z_1}\mathcal{A}=
\frac{\sqrt{Q_1-A_1}}{\Delta Q} \bigg [h -\beta \Delta A  \tanh (\beta h)-w_1\bigg].
\label{dAinz1}
\end{equation}
They read
\begin{equation}
    Z_t=\sqrt{Q_1-A_1} \bigg[ \frac{\chi_G}{\Delta Q}-\beta \frac{\Delta A}{\Delta Q}\chi_Q-\frac{m_1 W_t}{\Delta Q} \bigg], \qquad Z_h=\sqrt{Q_1-A_1} \bigg[ \frac{\chi_K}{\Delta Q}-\beta \frac{\Delta A}{\Delta Q}\chi_G-\frac{m_1 W_h}{\Delta Q} \bigg],
    \label{eqSM:ZtZh}
\end{equation}
whereas the third, through an exact Gaussian integration-by-parts, is
\begin{equation}
    \avmu{z_1w_1}= \sqrt{Q_1-A_1}+ \avmu{w_1 \, \partial_{z_1} \mathcal{A}} = \sqrt{Q_1-A_1} \bigg[ 1 +\frac{m_1}{\Delta Q} \bigg( W_h-\beta \Delta A W_t -\avmu{w_1^2} \bigg)\bigg].
    \label{eqSM:zw}
\end{equation}

Each of these carries an overall factor $\sqrt{Q_1-A_1}$ that
cancels the $1/\sqrt{Q_1-A_1}$ in Eq.~(\ref{eq:I2}) before any numerical division.
Adding $I_1$, the $\pm 1/\Delta Q$ terms cancel, and using the elementary
relations in Eqs.~(\ref{eqSM:ZtZh})--(\ref{eqSM:zw}), every surviving term acquires a common factor $(1-m_1)$:
\begin{equation}
 I_1 + I_2 = \frac{1-m_1}{\Delta Q^2}\Bigl[
 K_1 - 2 W_h + \avmu{w_1^2} - 2\beta\Delta A\, G_1 + 2\beta\Delta A\, W_t
 + \beta^2 \Delta A^2\, Q_1 \Bigr],
\end{equation}
so that the factor $(1-m_1)$ in Eq.~(\ref{eq:spC1raw}) cancels and the final 
expression for the saddle point equation for $C_1$ becomes
\begin{align}
 C_1 &= -2 m A_1 - m^2 Q_1 + \frac{\mathcal{N}_1}{\beta^2\,\Delta Q^2},
 \label{eq:C1stable} \\
 \mathcal{N}_1 &= K_1 - 2 W_h + \avmu{w_1^2} - 2\beta\Delta A\,(G_1 - W_t)
 + \beta^2 \Delta A^2 Q_1, \nonumber
\end{align}
as in Eq.~(\ref{EM:C1stable}).
The free energy of the states and the complexity, parametric in $m$, are
\begin{align}
 f^{\rm 1RSB}(m) =\;& -\frac{\avmu{\avh{\ln 2\cosh\beta h}}}{\beta}- \frac{\beta}{4}\bigl[(1-Q)^2 - 2 m Q^2 - 4 A Q\bigr]- \frac{\beta}{4}(1-m_1)\bigl[2 m Q_1^2 + 4 A_1 Q_1\bigr],
 \label{SM:fm}\\
 \Sigma^{\rm 1RSB}(m) =\;& \beta\phi_m^{\rm 1RSB}
 - m \avmu{\avh{\ln 2\cosh\beta h}}+ \frac{\beta^2}{4}\bigl[m^2 Q^2 - 4 A (1-Q) - 2(A^2+QC)\bigr] \nonumber\\
 &- \frac{\beta^2 (1-m_1)}{4}\bigl[m^2 Q_1^2 - 2(A_1^2 + Q_1 C_1)\bigr].
 \label{SM:Sigmam}
\end{align}
 In the limit $Q_1 = A_1 = C_1 = 0$, the saddle-point equations in Eqs.~(\ref{eq:spQ})--(\ref{eq:spC1raw}) and the expressions in Eqs.~(\ref{SM:fm}) and ~(\ref{SM:Sigmam})
collapse onto the annealed ones of Appendix~\ref{sec:annealed}. 
\section{$T \to 0$ limit of the quenched 1RSB theory}
\label{sec:T0}

At low temperature the diagonal parameters behave as in the annealed case,
$Q = 1 + w_Q T^2+ O(T^3)$, $A = k_A T + w_A T^2+O(T^3)$, $C = k_C T + w_C T^2+O(T^3)$, while the
inner-block parameters scale as
\begin{equation}
 Q_1 = l_{Q_1} + k_{Q_1} T + w_{Q_1}T^2+O(T^3), \qquad
 A_1 = k_{A_1}\,T + w_{A_1}\,T^2+O(T^3), \qquad
 C_1 = w_{C_1}\,T^2+O(T^3),
\end{equation}
as verified numerically against the finite-temperature solution at fixed
$y=\beta m$ for $T\to 0$. All the dependence on
$w_Q,w_A,w_C,k_C,k_{Q_1},w_{Q_1}$ and $w_{A_1}$ cancels, so that the $T=0$ theory
depends only on $l_{Q_1}$, $k_A$, $k_{A_1}$, $w_{C_1}$ and $m_1$, and the three
Gaussian fields $(z_1,t_1,y_1)$ reduce to two, namely
\begin{equation}
 w_1=\sqrt{l_{Q_1}}\,z_1,\qquad
 \lim_{\beta \to \infty} \beta \, v_1 = \tilde v_1
 =\frac{k_{A_1}}{\sqrt{l_{Q_1}}}\,z_1
 +\sqrt{w_{C_1}-\frac{k_{A_1}^2}{l_{Q_1}}}\;t_1 ,
 \label{eq:T0fields}
\end{equation}
where $(z_1,t_1)$ are independent standard Gaussian variables. Therefore, the $T=0$ replicated free energy is
\begin{align}
 -y\,\Phi^{\rm 1RSB,\,0}_{\rm 2G}(y) =\;&
 -\frac{k_A^2}{2} - y\,k_A - \frac{y^2}{4}
 - \frac{(k_A-k_{A_1})^2}{2\,(1-l_{Q_1})} - \frac{w_{C_1}}{2} + \frac{1-m_1}{2}\Bigl[k_{A_1}^2 + l_{Q_1}w_{C_1} + 2y\,l_{Q_1}k_{A_1}
 + \frac{y^2 l_{Q_1}^2}{2}\Bigr] \nonumber\\
 &+ (\beta\phi_m^{\rm 1RSB})^0 ,
 \label{eq:T0Phi}
\end{align}
with
\begin{align}
 (\beta\phi_m^{\rm 1RSB})^0 =\;& \frac1{m_1}\ln\int \mathcal{D}z_1\,\mathcal{D}t_1\,
 \bigl[\zeta_h^0(z_1,t_1)\bigr]^{m_1},\\
 \zeta_h^0(z_1,t_1) =\;& \int\frac{dh}{\sqrt{2\pi\,(1-l_{Q_1})}}\;
 e^{\mathcal{A}^0(h,z_1,t_1)},\\
 \mathcal A^0 (h, z_1, t_1)=\;& -\frac{\bigl(h-\sqrt{l_{Q_1}}\,z_1\bigr)^2}{2\,(1-l_{Q_1})}
 + \frac{k_A-k_{A_1}}{1-l_{Q_1}}\Bigl(|h|-\sqrt{l_{Q_1}}\,z_1\sign h\Bigr)
 + y|h| + \tilde v_1\sign h .
\end{align}

As in the annealed case, the integral over $h$ is a combination of Gaussian
integrals truncated to the two half-lines. For $h>0$ and $h<0$ the action
$\mathcal A^0$ is a Gaussian in $h$ with, respectively,
\begin{equation}
 u_\pm(z_1) \equiv \frac{\pm\sqrt{l_{Q_1}}\,z_1 + k_A - k_{A_1} + y\,(1-l_{Q_1})}
 {\sqrt{1-l_{Q_1}}} ,
 \label{eq:T0upm}
\end{equation}
playing the role of the parameter $u=y+k_A$ of Eq.~(\ref{eq:udefM}), which is
recovered for $l_{Q_1}=k_{A_1}=0$ and $z_1=0$. With the same function
\begin{equation}
 M(u)\equiv\sqrt{\frac{2}{\pi}}\,\frac{e^{-u^{2}/2}}{{\rm Erfc}(-u/\sqrt2)}
\end{equation}
of Eq.~(\ref{eq:udefM}), one finds $\zeta_h^0=Z_++Z_-$ with
\begin{equation}
 Z_\pm(z_1,t_1)=\frac12\,
 \exp\!\Bigl\{-\frac{l_{Q_1}z_1^{2}}{2(1-l_{Q_1})}
 \mp\frac{(k_A-k_{A_1})\sqrt{l_{Q_1}}\,z_1}{1-l_{Q_1}}\pm\tilde v_1\Bigr\}\,
 e^{u_\pm^{2}/2}\,{\rm Erfc}\left(-\frac{u_\pm}{\sqrt2}\right),
 \label{eq:T0Zpm}
\end{equation}
where $\tilde v_1$ is given by Eq.~(\ref{eq:T0fields}) and $u_\pm=u_\pm(z_1)$. The two half-lines carry the normalized weights
\begin{equation}
 p_\pm \equiv \frac{Z_\pm}{Z_++Z_-},
 \qquad p_++p_-=1 ,
\end{equation}
and, since on each half-line $h/\sqrt{1-l_{Q_1}}$ is a Gaussian truncated to
positive (negative) values with parameter $u_\pm$, all averages over $h$,
\begin{equation}
 \avh{X} = \frac{1}{\zeta_h^0}\int\frac{dh}{\sqrt{2\pi\,(1-l_{Q_1})}}\,
 e^{\mathcal{A}^0}\,X ,
\end{equation}
are explicit here
\begin{align}
 \avh{\sign h} &= p_+ - p_- ,\nonumber\\
 \avh{|h|} &= \sqrt{1-l_{Q_1}}\,\Bigl\{p_+\bigl[u_+ + M(u_+)\bigr]
 + p_-\bigl[u_- + M(u_-)\bigr]\Bigr\},
 \label{eq:T0hmoments}\\
 \avh{h} &= \sqrt{1-l_{Q_1}}\,\Bigl\{p_+\bigl[u_+ + M(u_+)\bigr]
 - p_-\bigl[u_- + M(u_-)\bigr]\Bigr\}.\nonumber
\end{align}
For $l_{Q_1}=k_{A_1}=w_{C_1}=0$ one has $u_+=u_-=u=y+k_A$, $\tilde v_1=0$,
$p_\pm=\tfrac12$ and the above reduce to
$\zeta_h^0=e^{u^{2}/2}\,{\rm Erfc}(-u/\sqrt2)$ and $\avh{|h|}=u+M(u)$, i.e.\ to
Eqs.~(\ref{SM:bigPhi0}) and (\ref{eq:momentsT0}) of the annealed theory, and
Eq.~(\ref{eq:T0Phi}) reduces to Eq.~(\ref{SM:bigPhi0}).

The outer average is
\begin{equation}
 \avmu{Y} = \frac{1}{\zeta_{\mu}}\int\mathcal{D}z_1\,\mathcal{D}t_1\,
 [\zeta_h^0]^{m_1}\,Y, \qquad
 \zeta_{\mu}=\int\mathcal{D}z_1\,\mathcal{D}t_1\,[\zeta_h^0]^{m_1},
\end{equation}
while the averages over $h$ are those of Eq.~(\ref{eq:T0hmoments}).
Stationarity of Eq.~(\ref{eq:T0Phi}) with respect to $w_{C_1},k_A,k_{A_1},l_{Q_1}$ gives the four saddle point equations
\begin{align}
 l_{Q_1} =\;& \avmu{\avh{\sign h}^2},
 \label{eq:T0lq}\\
 k_A =\;& \frac{1}{1-l_{Q_1}}\Bigl[\avmu{\avh{|h|}}
 - \sqrt{l_{Q_1}}\,\avmu{z_1\avh{\sign h}} - (k_A-k_{A_1})\Bigr]-y,
 \label{eq:T0kA}\\
 k_{A_1} =\;& -y\,l_{Q_1} + \frac{1}{1-l_{Q_1}}
 \Bigl[\avmu{\avh{h}\avh{\sign h}}
 - \sqrt{l_{Q_1}}\,\avmu{z_1\avh{\sign h}}
 - (k_A-k_{A_1})\avmu{\avh{\sign h}^2}\Bigr],
 \label{eq:T0kA1}\\
 w_{C_1} =\;& -2y\,k_{A_1} - y^2 l_{Q_1} + \frac{1}{(1-l_{Q_1})^2}
 \Bigl[\avmu{\avh{h}^2}
 - 2\sqrt{l_{Q_1}}\,\avmu{z_1\avh{h}}
 + l_{Q_1}\avmu{z_1^2} \nonumber\\
 &- 2(k_A-k_{A_1})\avmu{\avh{h}\avh{\sign h}}
 + 2(k_A-k_{A_1})\sqrt{l_{Q_1}}\,\avmu{z_1\avh{\sign h}}
 + (k_A-k_{A_1})^2\avmu{\avh{\sign h}^2}\Bigr],
 \label{eq:T0wC1}
\end{align}
and the one for the parameter $m_1$ reads
\begin{equation}
\avmu{\ln\zeta_h^0} - (\beta\phi_m^{\rm 1RSB})^0 =\;
 \frac{m_1}{2}\Bigl[k_{A_1}^2 + l_{Q_1}w_{C_1} + 2y\,l_{Q_1}k_{A_1}
 + \frac{y^2 l_{Q_1}^2}{2}\Bigr].
 \label{eq:T0m1}
\end{equation}
with $\zeta_h^0=Z_++Z_-$ from Eq.~(\ref{eq:T0Zpm}).
Eqs.~(\ref{eq:T0lq})--(\ref{eq:T0wC1}) are the limits of Eqs.~(\ref{eq:spQ1})--(\ref{eq:spA1}) and (\ref{eq:C1stable}), written in terms of the
direct averages rather than of the integrated-by-parts combinations used at finite
$T$, and Eq.~(\ref{eq:T0m1}) is the limit of Eq.~(\ref{eq:spm1}).

The energy and the ground-state complexity as functions of $y=\beta m$ are
\begin{align}
 e(y) =\;& k_A - \avmu{\avh{|h|}} + \frac y2
 - (1-m_1)\Bigl(l_{Q_1}k_{A_1} + \frac{y\,l_{Q_1}^2}{2}\Bigr),\\
 \Sigma(y)=\;& -\frac{k_A^2}{2} - \frac{(k_A-k_{A_1})^2}{2\,(1-l_{Q_1})} - \frac{w_{C_1}}{2}
 - y\,\avmu{\avh{|h|}} + \frac{y^2}{4}+ \frac{1-m_1}{2}\Bigl[k_{A_1}^2 + l_{Q_1}w_{C_1}
 - \frac{y^2 l_{Q_1}^2}{2}\Bigr] + (\beta\phi_m^{\rm 1RSB})^0.
\end{align}

\section{Reduced Ansatz and Comparison}
\label{sec:reduced}
The seven saddle point equations in Eqs.~(\ref{eq:spQ})--(\ref{eq:spC}), (\ref{eq:C1stable}) and (\ref{eq:spm1}) are solved through the numerical methods in Appendix~\ref{sec:numerics}. The converged resulting values for $A_1$ and $C_1$ are particularly interesting: over the whole range of temperatures explored and in the entire range of $m$, $|A_1|$ ranges
between about $10^{-4}$ and $10^{-3}$, while $A$ ranges between about $10^{-3}$ and $10^{-2}$; $C_1$ ranges between about $10^{-7}$ and $10^{-5}$, while $C$ ranges between about $10^{-2}$ and $10^{-1}$. This might motivate the introduction of a \textit{reduced} ansatz, in which we impose $A_1=C_1=0$ from the beginning of the computation of the two-group replicated free energy $\Phi_{2G}$.  Physically, this says that different marginal states belonging
to the same cluster have {\em uncorrelated} soft-mode directions, while their
magnetizations remain correlated through $Q_1\neq 0$.

\begin{figure}[t!]
    \centering
    \includegraphics[width=0.31\linewidth]{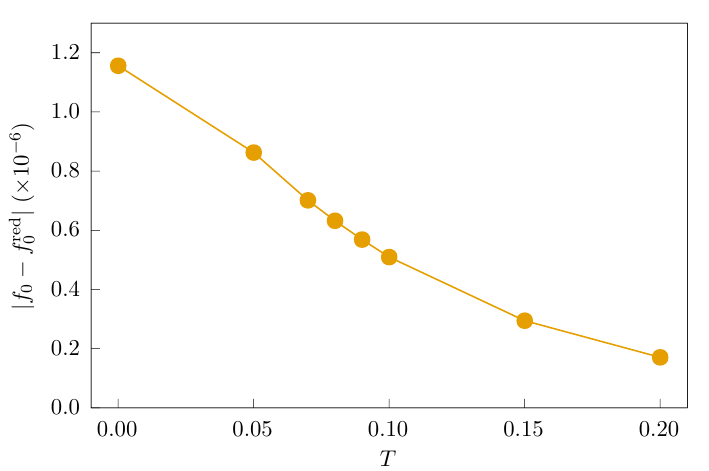}
     \includegraphics[width=0.31\linewidth]{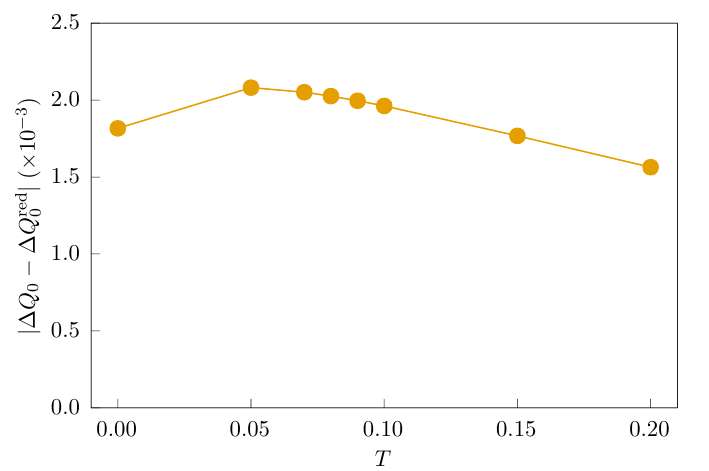}
     \includegraphics[width=0.31\linewidth]{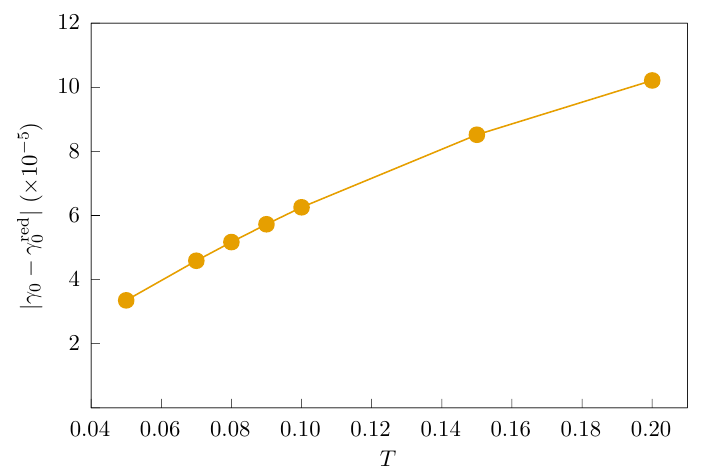}
   \caption{Left: absolute difference $|f_0-f_0^{\rm red}|$ (in units of $10^{-6}$) between the \textit{complete} and the \textit{reduced} ansatz as a function of the temperature $T$. The difference is negative for all $T$ and decreases monotonically with increasing temperature. Middle:  $|\Delta Q_0-\Delta Q_0^{\rm red}|$ (in units of $10^{-3}$): the difference is negative for all $T$.
 Right: Absolute difference $|\gamma_0-\gamma_0^{\rm red}|$
(in units of $10^{-5}$). The difference is negative for all $T$ and increases with temperature. }
    \label{fig:diff}
\end{figure}

\begin{figure}[t!]
    \centering
    \includegraphics[width=0.5\linewidth]{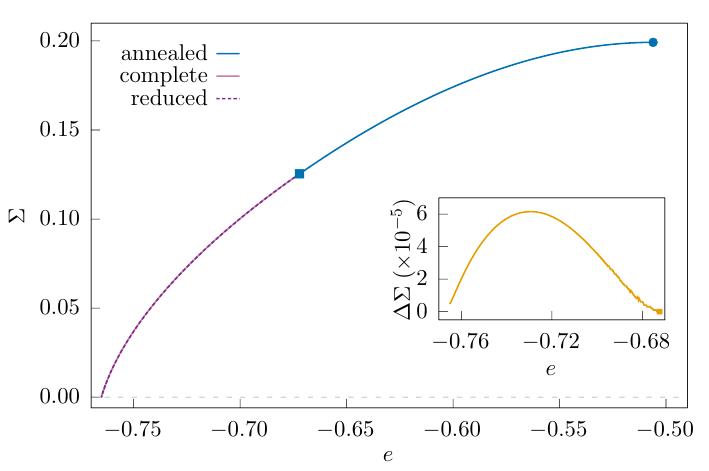}
    \caption{Complexity $\Sigma$ versus energy density $e$ at $T=0$. The annealed branch (blue) at its instability point exactly continues in the quenched branch, evaluated both in the \textit{complete} ansatz (pink), the one used in the main text, and in the \textit{reduced} ansatz (dotted purple). The two seem to coincide up to order $10^{-5}$: in the inset their difference versus the energy density $e$. }
    \label{fig:diff_sigma}
\end{figure}

\begin{table}[t!]
 \small
 \setlength{\tabcolsep}{2pt}
 \begin{ruledtabular}
 \begin{tabular}{ccccccc}
  $T$ & Method & $y_0$ & $f_0$ & $\Delta Q_0$ & $\gamma_0$ & $s_0$ \\
 \hline
 0.00
  & Complete & 4.01330096717 & -0.765203540763 & 0.1247234066 & -- & -- \\
  & Reduced  & 4.00990125281 & -0.765202384733 & 0.1265395947 & -- & -- \\
 \hline
 0.05
  & Complete & 2.6874422342 & -0.764288924923 & 0.1898897777 & 1.5296434959 & -0.0100823706 \\
  & Reduced  & 2.6854282890 & -0.764288062253 & 0.1919706986 & 1.5296770232 & -0.0100739354 \\
 \hline
 0.07
  & Complete & 2.3970553752 & -0.764133334476 & 0.2115479394 & 1.5226091348 & -0.0055614315 \\
  & Reduced  & 2.3954280090 & -0.764132633301 & 0.2135990057 & 1.5226550217 & -0.0055540982 \\
 \hline
 0.08
  & Complete & 2.2768980341 & -0.764088420116 & 0.2215596190 & 1.5196786465 & -0.0034279764 \\
  & Reduced  & 2.2754289944 & -0.764087787936 & 0.2235853416 & 1.5197303499 & -0.0034212460 \\
 \hline
 0.09
  & Complete & 2.1695210910 & -0.764064713790 & 0.2310521257 & 1.5170510187 & -0.0013133704 \\
  & Reduced  & 2.1681911581 & -0.764064145167 & 0.2330478058 & 1.5171082774 & -0.0013072498 \\
 \hline
 0.10
  & Complete & 2.0728126197 & -0.764062198235 & 0.2400424328 & 1.5146809372 & 0.0008146182 \\
  & Reduced  & 2.0716054507 & -0.764061688834 & 0.2420045176 & 1.5147434955 & 0.0008201451 \\
 \hline
 0.15
  & Complete & 1.7013430263 & -0.764382748908 & 0.2779307666 & 1.5056821045 & 0.0123089989 \\
  & Reduced  & 1.7005684818 & -0.764382454422 & 0.2796982807 & 1.5057672892 & 0.0123122316 \\
 \hline
 0.20
  & Complete & 1.4450611640 & -0.765331009609 & 0.3049535140 & 1.4999017003 & 0.0260458813 \\
  & Reduced  & 1.4445343758 & -0.765330839013 & 0.3065172081 & 1.5000038253 & 0.0260477475 \\
 \end{tabular}
 \end{ruledtabular}
 \caption{Zero-complexity quantities ($\Sigma=0$) as a function of $T$ for the complete and reduced ansatz
 ($f_0=e_0$, $\Delta Q_0=1-Q_1$). At $T=0$ the quantities $\gamma_0$ and $s_0$ are not reported. 
 The mass $m_0$ can be recovered from $m_0 = y_0\,T$.}
 \label{tab:zero_complexity}
\end{table}

A comparison between the results obtained from the two methods to which refer to as \textit{complete} and \textit{reduced} ansatz is shown in Tab.~\ref{tab:zero_complexity}, which underlines an agreement up to about $10^{-6}$ for the zero-complexity free
energy $f_0$, $10^{-5}$ for $s_0$, and $10^{-4}$ for  $m_1$ and $\gamma$, while the largest differences, of order $10^{-3}$, are found for $y$ and $\Delta Q$. Fig.~\ref{fig:diff} show, as a
function of $T$, some examples of the difference between the values obtained with the \textit{complete} and with the \textit{reduced} ansatz of, respectively, the zero complexity energy density $f_0$, difference between the overlap $Q$ and $Q_1$ and the angle between the magnetization and the soft mode $\gamma$. Finally, Fig.~(\ref{fig:diff_sigma}) displays the complexity $\Sigma$ as a function of the energy density $e$ at $T=0$, evaluated in the 1RSB branch with both the ansatzes. The curves coincide up to a difference, as demonstrated in the inset, of order $10^{-5}$. 


\section{Numerical methods}
\label{sec:numerics}

\subsection{Finite temperature}

\paragraph*{Saddle-point equations.}
At fixed $(T,m,m_1)$ the six overlaps $(Q,Q_1,A,A_1,C,C_1)$ are a stationary
point of $-\beta m\Phi_{2G}^{\rm 1RSB}$. The kernel (\ref{eq:kernel}) depends on
$\mu_1$ only through the composed fields $(w_1,v_1)$, jointly Gaussian with
covariance
$\Sigma=\bigl(\begin{smallmatrix}Q_1&A_1\\A_1&C_1\end{smallmatrix}\bigr)$.
The derivatives of this Gaussian measure with respect to the inner-block
overlaps follow from
\begin{equation}
 \partial_{Q_1}\mathbb E[F]=\tfrac12\mathbb E[\partial^2_{w_1}F],\qquad
 \partial_{C_1}\mathbb E[F]=\tfrac12\mathbb E[\partial^2_{v_1}F],\qquad
 \partial_{A_1}\mathbb E[F]=\mathbb E[\partial_{w_1}\partial_{v_1}F],
 \qquad F\equiv\zeta_h^{m_1}.
\end{equation}
With $D\equiv h-w_1-\beta\Delta A\tanh\beta h$, the saddle-point equations
take the form
\begin{align}
 1-Q &= \avmu{\avh{1-\tanh^2\beta h}}, \qquad
 Q_1 = \avmu{\avh{\tanh\beta h}^2}, \nonumber\\
 \beta(1-Q)+\beta(A+mQ) &= \frac{G-W_t-\beta\Delta A\,Q}{\Delta Q}, \qquad
 A_1 = -mQ_1 + \frac{\avmu{\avh{D}\avh{\tanh\beta h}}}{\beta\Delta Q}, \nonumber\\
 C &= 2A(1-m)-m(1-Q)-m^2Q+\frac{\avmu{\avh{D^2}}-\Delta Q}{\beta^2\Delta Q^2}, \nonumber\\
 C_1 &= -2mA_1-m^2Q_1+\frac{\avmu{\avh{D}^2}}{\beta^2\Delta Q^2}.
 \label{eq:SPnum}
\end{align}
By Gaussian integration by parts they are equivalent to
Eqs.~(\ref{eq:spQ})--(\ref{eq:spC}) and (\ref{eq:C1stable}), with
$\avmu{\avh{D}^2}=\mathcal N_1$ and
$(\avmu{\avh{D^2}}-\Delta Q)/\Delta Q^2=I_1$. They contain no square roots of
$Q_1-A_1$, $C_1-A_1$ or $A_1$, hence hold for any real $\Sigma$: the only
constraint is $0<Q_1<Q$.

\paragraph*{Outer fields and indefinite covariance.}
The outer average is generated from two independent standard normal variables
$a,b$,
\begin{equation}
 w_1=\sqrt{Q_1}\,a,\qquad v_1=\frac{A_1}{\sqrt{Q_1}}\,a+s\,b,\qquad
 s^2\equiv C_1-\frac{A_1^2}{Q_1},
 \label{eq:map}
\end{equation}
which, unlike the representation in $(z_1,t_1,y_1)$, requires neither
$A_1\ge0$ nor $C_1\ge A_1$. At the solutions, $A_1<0$ and $s^2>0$ is tiny (from
$3\times10^{-9}$ at $T=0.005$ to $6\times10^{-7}$ at $T=0.1$, scaling as $T^2$),
so Newton iterates can reach $s^2<0$. Since the integrand depends on $v_1$ only
through $e^{\beta v_1\tanh\beta h}$, the $b$ average is an analytic function of
$s^2$ and is continued to $s^2<0$ by $s\to i|s|$. With a symmetric
Gauss--Hermite rule all averages stay real. No clamp on $A_1$ or $C_1$ is
needed. A clamp $A_1\ge0$ would exclude the true saddle point.

\paragraph*{Quadrature.}
The $h$ integral uses composite Gauss--Legendre quadrature ($n_h$ nodes per
panel) on each half-line: geometric panels from width $1/(4\beta)$, doubling
up to $\sqrt{\Delta Q}$, resolve the boundary layer of width $T$ at $h=0$ at a
cost growing like $\ln\beta$. Uniform panels of width $\sqrt{\Delta Q}$ then
extend $10\sqrt{\Delta Q}$ beyond the lobe centres
$h\simeq w_1\pm\Delta Q\,\kappa$, with $\kappa\equiv\beta m+\beta\Delta A/\Delta Q$.

By the symmetry $(h,a,b)\to(-h,-a,-b)$ the $a$ integral is folded onto
$a\ge0$ and computed by composite Gauss--Legendre quadrature ($n_a$ nodes per
panel, geometric panels from $0.05$ to $0.5$), truncated at
$a_{\max}=m_1\kappa\sqrt{Q_1}+11$, i.e.\ 11 standard deviations beyond the
centre of the reweighted measure. 

The $b$ integral uses $n_b$ Gauss--Hermite
nodes. 

With the  values $(n_h,n_a,n_b)=(10,8,8)$, used to produce the results in this manuscript, all reported
quantities differ by less than $10^{-12}$ from those obtained with a more refined grid 
$(20,18,28)$.

\paragraph*{Reweighting.}
With $\ell_k=\ln\zeta_h^{(k)}$ on the outer nodes (weights $\omega_k$,
$\sum_k\omega_k=1$) and $\bar\ell=\sum_k\omega_k\ell_k$,
\begin{equation}
 \beta\phi=\bar\ell+\frac1{m_1}\ln(1+\delta),\qquad
 \delta=\sum_k\omega_k\bigl[e^{m_1(\ell_k-\bar\ell)}-1\bigr],
\end{equation}
where $\delta$ and $\ln(1+\delta)$ are evaluated without cancellation, so that
the accuracy does not degrade as $m_1\to0$. The reweighted averages are
\begin{equation}
 \avmu{X}=\frac{\sum_k\omega_k\,e^{m_1\ell_k-M}X_k}{\sum_k\omega_k\,e^{m_1\ell_k-M}},
 \qquad M=\max_k m_1\ell_k ,
\end{equation}
which prevents overflow at low $T$, where $\ell_k=O(\beta)$.

\paragraph*{Newton solver.}
In the rescaled unknowns
\begin{equation}
 x=\bigl(\beta^2(1-Q),\;Q_1,\;\beta A,\;\beta A_1,\;\beta C,\;\beta C_1\bigr),
\end{equation}
which remain $O(1)$ at all temperatures, Eqs.~(\ref{eq:SPnum}) are written as
$R(x)=0$, the $Q_1$ residual being divided by $Q_1$ to remove the trivial root
$Q_1=0$. Each step solves $J\Delta x=-R(x)$ and accepts
$x+\lambda\Delta x$ at the first $\lambda=1,\frac12,\frac14,\dots$ such that
\begin{equation}
 \|R(x+\lambda\Delta x)\|<(1-10^{-4}\lambda)\,\|R(x)\| .
\end{equation}
The Jacobian is computed by forward differences at the first step and then
updated, with $\sigma=\lambda\Delta x$ and $r=R(x+\sigma)-R(x)$, as
\begin{equation}
 J\;\to\;J+\frac{(r-J\sigma)\,\sigma^{\top}}{\sigma^{\top}\sigma},
\end{equation}
which needs no new evaluation of the averages; it is recomputed by finite
differences whenever the step-length search fails. Convergence is declared at
$\max_i|R_i|<10^{-11}$. Solutions with $Q_1\to0$ or $A=C=0$ (the one-group
root) are rejected. At the solutions, the six partial derivatives of
$-\beta m\Phi_{2G}^{\rm 1RSB}$, computed by centred finite differences,
vanish within the differencing accuracy.

\paragraph*{Replica symmetry breaking parameter $m_1$ and continuation in $m$.}
At each $m$, $m_1^*=\arg\min_{m_1}(-\beta m\Phi_{2G}^{\rm 1RSB})$, with the
overlaps re-solved at each trial $m_1$: the minimum is bracketed by a
seven-point scan around the previous $m_1^*$ and refined by parabolic
interpolation  to relative accuracy
$10^{-7}$. The stationarity condition (\ref{eq:spm1}) then holds at $m_1^*$.
The annealed branch gives $m^*(T)$ by bisection on the zero of the replicon;
the 1RSB branch is followed in steps $\Delta m=0.005$ with warm starts until
$\Sigma$ changes sign. 

The zero $\Sigma(m_0)=0$ is then located by linear interpolation between the
two bracketing points, re-optimizing $m_1$ at each new point; whenever the
same bracketing end is kept, its value of $\Sigma$ is halved before the next
interpolation, and the iteration stops at $|\Sigma|<10^{-12}$.

The quantities $f$ and $\Sigma$ follow from
Eqs.~(\ref{SM:fm})--(\ref{SM:Sigmam}). The
cost per solution is nearly independent of $\beta$, allowing temperatures
down to $T=0.005$.

\subsection{Zero temperature}
At fixed $y=\beta m$, with
\begin{equation}
 Q=1-uT^2,\quad A=k_AT,\quad C=k_CT,\quad Q_1\to l_{Q_1},\quad
 A_1=k_{A_1}T,\quad C_1=w_{C_1}T^2
\end{equation}
(these scalings are checked with the finite-$T$ solver down to $T=0.005$), the
functional depends only on $(l_{Q_1},k_A,k_{A_1},w_{C_1})$ and $m_1$. The outer
pair $(w_1,\tilde v_1=\beta v_1)$ is Gaussian with covariance
$\bigl(\begin{smallmatrix}l_{Q_1}&k_{A_1}\\k_{A_1}&w_{C_1}\end{smallmatrix}\bigr)$.
Since $\tanh\beta h\to\sign h$, the inner integral is
\begin{equation}
 \zeta_h^0=\frac{e^{\Delta a^2/2}}{2}\Bigl[e^{\,yw_1+\tilde v_1}\,
 \operatorname{erfc}\Bigl(-\frac{w_1+\Delta a}{\sqrt{2\Delta}}\Bigr)
 +e^{-yw_1-\tilde v_1}\,\operatorname{erfc}\Bigl(\frac{w_1-\Delta a}{\sqrt{2\Delta}}\Bigr)\Bigr],
 \qquad \Delta=1-l_{Q_1},\quad a=y+\frac{k_A-k_{A_1}}{\Delta},
\end{equation}
and the inner averages are moments of Gaussians restricted to each half-line,
evaluated through asymptotic expansions of $\operatorname{erfc}$ for large
arguments. The outer average is computed as at finite $T$, with the map
(\ref{eq:map}). With $D_0=h-w_1-(k_A-k_{A_1})\sign h$, the stationarity
conditions are
\begin{align}
 l_{Q_1}&=\avmu{\avh{\sign h}^2},\qquad
 k_A+y=\frac{\avmu{\avh{|h|}}-\avmu{w_1\avh{\sign h}}-(k_A-k_{A_1})}{\Delta},
 \nonumber\\
 k_{A_1}&=-y\,l_{Q_1}+\frac{\avmu{\avh{D_0}\avh{\sign h}}}{\Delta},\qquad
 w_{C_1}=-2yk_{A_1}-y^2l_{Q_1}+\frac{\avmu{\avh{D_0}^2}}{\Delta^2},
\end{align}
solved by the same Newton iteration (Jacobian recomputed at every step,
tolerance $10^{-12}$), with $m_1^*$ optimized at each $y$ and the zero
$\Sigma(y_0)=0$ located as at finite $T$, to $|\Sigma|<10^{-13}$. At the
solutions $k_{A_1}<0$ and $0<w_{C_1}-k_{A_1}^2/l_{Q_1}<2\times10^{-4}$. The
ground-state energy of the dominant marginal states is $e_0=-0.765203543$. We notice also that
setting $k_{A_1}=w_{C_1}=0$ -- see the reduced Ansatz of Appendix~\ref{sec:reduced}  --  changes $e_0$ only by $10^{-6}$.


\section{The angle between magnetization and soft mode}
\label{app:EMgamma}

The two-group construction fixes the {\em direction} of the soft mode
$\delta\tilde{\bm m}$ of a marginal state, but not its magnitude. The
minimum--saddle pair enters only through
$\widetilde m^{\pm}_i=\widetilde m_i\pm\delta\widetilde m_i/2K$
\cite{Mueller06}, so that $\delta\tilde{\bm m}$ and the group size $K$ appear
in Eq.~(\ref{eq:2Gansatz}), solely through the ratio $\delta\tilde{\bm m}/K$.
Since $K$ is a formal parameter, sent to infinity, the splitting of that ratio
into the length of $\delta\tilde{\bm m}$ and the value of $K$ is arbitrary:
the combined transformation
$\delta\tilde{\bm m}\to\lambda\,\delta\tilde{\bm m}$, $K\to\lambda K$ leaves
$\widetilde m^{\pm}_i$, and hence every overlap, unchanged, while
$A_{ab}\to\lambda A_{ab}$ and $C_{ab}\to\lambda^{2}C_{ab}$.
Normalizing the soft modes as
$\delta\widehat{\widetilde m}_i=\delta\widetilde m_i/\sqrt{C_{aa}}$
\cite{Mueller06}, the generic angle in the $N$-dimensional space between the TAP magnetization of a state $a$ and the soft mode of a state $b$ is defined as
\begin{equation}
 \cos\gamma_{ab}
 = \frac{\langle \delta\widetilde m^a_i\,\widetilde m^b_i\rangle}
 {\sqrt{\langle \delta\widetilde m_i^2\rangle\langle \widetilde m_i^2\rangle}}
 = \frac{A_{ab}}{\sqrt{Q_{aa}C_{aa}}},
 \label{EM:gammaab}
\end{equation}
namely a combination of the overlap matrices that survives the rescaling.
Eq.~(\ref{eq:gamma}) is its diagonal element $\gamma\equiv\gamma_{aa}$, which indeed depends only on the diagonal values of the overlap matrices, namely on the intra-state correlations between minima and saddles.
For $\gamma\to0$ the soft mode is radial, aligned with $\tilde{\bm m}$: motion along the flat direction changes $|\tilde{\bm m}|^{2}$, that is the self-overlap $Q$, and carries
the system out of the state. On the other hand, for $\gamma\to\pi/2$ the soft mode is transverse
and $Q$ is unchanged to leading order. Marginality alone does not, therefore, make a state
easy to leave, and $\gamma$ measures how much of it is usable as an escape
route. If the slow relaxation proceeds along the soft mode, the rate at which
the Edwards--Anderson parameter decreases is governed by $\cos\gamma$ and
slows down as $\gamma\to\pi/2$ \cite{Mueller06}.

Along the quenched branch $\gamma$ grows towards $\pi/2$, and at the
zero-complexity endpoint it lies closer to orthogonality than in the annealed
theory at every temperature, see Tab.~\ref{tab:gamma0} and the top panel of Fig.~\ref{fig:gamma_allT}.
\begin{table}[t!]
 \begin{ruledtabular}
 \begin{tabular}{cccc}
 $T$ & $\gamma_0^{\rm ann}$ & $\gamma_0^{\rm 1RSB}$
 & $\pi/2-\gamma_0^{\rm 1RSB}$ \\
 \hline
 0.05 & 1.50435 & 1.5296 & 0.0412 \\
 0.10 & 1.47813 & 1.5147 & 0.0561 \\
 0.20 & 1.45017 & 1.4999 & 0.0709 \\
 \end{tabular}
 \end{ruledtabular}
 \caption{Angle $\gamma$ between the magnetization of a marginal state and
 its soft mode, Eq.~(\ref{eq:gamma}), evaluated at the zero-complexity
 endpoint, in the annealed and in the quenched 1RSB two-group computations.
 The quenched value is closer to orthogonality, $\gamma=\pi/2=1.57080$, at
 every temperature, and the residual deficit shrinks as $T$ decreases.}
 \label{tab:gamma0}
\end{table}
That $\pi/2$ is the equilibrium value follows from the nature of
FRSB marginality: equilibrium states are marginal because the replicon is
massless, but their flat directions are transverse. A soft mode with a
component along $\tilde{\bm m}$ would make the magnitude of the magnetization
itself unstable, which is incompatible with   $f_0^{\rm eq}$ being the lowest level of the free energy landscape.
The approach $\gamma\to\pi/2$ is therefore the geometric counterpart, at the
level of the order parameters, of the statement that the lowest-lying marginal
states are the equilibrium ones.

\section{Entropy of the two-group 1RSB solution}
\label{sec:entropy}
The quenched 2G 1RSB solution, though improving much on the annealed solution, is still an unstable approximation. Though we do not provide a stability analysis,
an internal check on the quenched 1RSB solution is provided by the
quenched 1RSB 2G entropy, computed from the 2G free energy Eq.~(\ref{eq:Phi1RSB}), as
\begin{equation}
 s^{\rm 2G} = \beta^2\frac{\partial \Phi_{\rm 2G}}{\partial \beta}  .
\end{equation}
That is, the \textit{thermodynamic entropy}, whose reference \textit{thermodynamic free-energy} is $\Phi_m^{2G}$.
Since the model is Ising, thermodynamic consistency requires $s_{\rm 2G}\geq 0$, zero being available as $T\to 0$. A negative value of $s_{\rm 2G}$ is the classical signature of an inconsistency of the saddle point assumptions made to carry out the computation, that is, an insufficient level of replica symmetry breaking \cite{Parisi80,deAlmeida78}.
Analyzing the temperature behavior of the entropy density
at the zero-complexity point (that is the candidate thermodynamic equilibrium solution), cf.\ Table~\ref{tab:zero-compl_values}, it is possible to argue that for $T\simeq 0.095$  it turns negative. This is not unexpected: the one-step (1RSB) scheme adopted here for the structure of the marginal states is an approximation, and this is where its approximate nature becomes manifest. The small negative entropy is an independent indication that the
exact treatment requires the full replica symmetry breaking Ansatz among
the marginal states, in the direction conjectured in
Eq.~(\ref{eq:interpolation}).

\begin{figure}
    \centering
    \includegraphics[width=0.5\columnwidth]{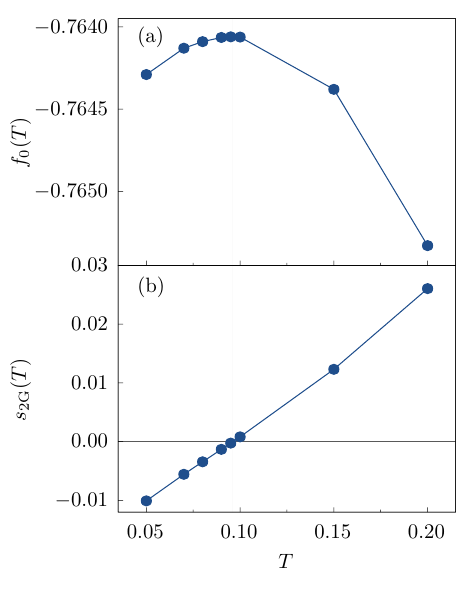}
    \caption{Thermodynamic consistency of the two-group 1RSB solution at the zero-complexity point. (a)~Zero-complexity free energy $f_0$ and (b)~2G entropy density $s_0$, computed in the 1RSB scheme, versus the
    temperature $T$. The $s_{\rm 2G}=0$ crossing ($T\simeq0.095$) corresponds to the maximum of $f_0(T)$; their coincidence reflects the identity $s_{\rm 2G}=-\,df_0/dT$, so that on lowering $T$ the free energy increases while $s_{\rm 2G}>0$ and decreases once $s_{\rm 2G}<0$.}
    \label{fig:entropy}
\end{figure}

Looking at the values of the zero complexity and the equilibrium free energies in Tab.~\ref{tab:zero-compl_values}, we just point out a possibly meaningful correlation between the $f_0(T)$ and the $s_{\rm 2G}(T)$ behaviors, as shown in Fig.~\ref{fig:entropy}.
Indeed, in the annealed case, in which $s_{\rm 2G}^{\rm ann}$ is negative at all considered $T\leq 0.2$, the zero complexity free energy decreases with $T$. At the opposite side, in the stable quenched full RSB equilibrium case, whose entropy (now a ``one group'' entropy at this specific equilibrium point, that is, the Parisi solution) is always positive (eventually zero at $T=0$), $f_{0}^{\rm eq}(T)$ increases with $T$.
The behavior of $f_0^{\rm 1RSB}(T)$ is, instead, non-monotonous and looking at the values of $s_{\rm 2G}(T)$ in Table~\ref{tab:zero-compl_values} we can observe that when $T$ decreases, $f_0^{\rm 1RSB}(T)$ increases as far as $s_{\rm 2G}(T)>0$ and decreases when $s_{\rm 2G}(T)<0$.

\begin{table}[b!]
 \caption{2G entropy and free energy values at zero complexity in the one step RSB scheme of computation.}
 \label{tab:zero-compl_values}
 \begin{ruledtabular}
 \begin{tabular}{cccccc}
 $T$ & $s^{2G}_0$-ann & $s_0^{2G}$-1RSB &$f_0$-ann & $f_0$-1RSB &$f_0$-FRSB \\
 \hline
 0.05 & -0.15533 & -0.01008 & -0.77979 & -0.76429 & -0.76333 \\
 0.07 & -0.12464 & -0.00556 & -0.77701 & -0.76413 & -0.76336\\
 0.08 & -0.11123 & -0.00343 & -0.77583 & -0.76409 & -0.76338\\
 0.09 & -0.09911 &  -0.00131 & -0.77478 & -0.76406 & -0.76341\\
 0.10 & -0.08793 &  0.00081 & -0.77384 & -0.76406 & -0.76345 \\
 0.15 & -0.04318 &  0.01231 & -0.77063 & -0.76438 & -0.76399\\
 0.20 & -0.00943 &  0.02605 & -0.76935 & -0.76533 & -0.76508 \\
 \end{tabular}
 \end{ruledtabular}
\end{table}

\begin{table}[b!]
 \caption{2G observables values at zero complexity in the one step RSB scheme of computation.}
 \label{tab:zero-compl_values_new}
 \begin{ruledtabular}
 \begin{tabular}{ccccccc}
 $T$ & $y_0$-ann & $y_0$-1RSB &$\gamma_0$-ann & $\gamma_0$-1RSB &$\Delta Q$-ann & $\Delta Q$-1RSB\\
 \hline
 0.05 & 1.14536 & 2.68744 & 1.50435 & 1.52960 & 0.99804 & 0.18989 \\
 0.07 & 1.10191 & 2.39706 & 1.49225 & 1.52266 & 0.99553 & 0.21360\\
 0.08 & 1.08077 & 2.27690 & 1.48709 & 1.51973 & 0.99379 & 0.22359\\
 0.09 & 1.06015 & 2.16952 & 1.48240 & 1.51718 & 0.99173 & 0.23105\\
 0.10 & 1.03997 & 2.07281 & 1.47813 & 1.51472 & 0.98933 & 0.24004 \\
 0.15 & 0.94602 & 1.70134 & 1.46143 & 1.50575 & 0.97222 & 0.27793\\
 0.20 & 0.86236 & 1.44506 & 1.45017 & 1.49991 & 0.94664 & 0.30495 \\
 \end{tabular}
 \end{ruledtabular}
\end{table}

\section{On the notion of marginality}
\label{sec:marginality_notions}
Several notions of marginality circulate in the literature on random
landscapes, and it is worth stating explicitly which one underlies the present
computation. Here, as in Refs.~\cite{Bray81,Aspelmeier04,Mueller06}, a state is
marginal when its TAP free-energy Hessian possesses a \emph{single} soft mode,
whose \emph{isolated} eigenvalue vanishes in the thermodynamic limit while the
spectral density at the origin remains zero. This is a property of the
\emph{typical, dominant} TAP states of the SK model at all free energies below
$T_c$, and it is exactly what forces the BRST-breaking or two-group counting
adopted here: the two groups of replicas describe the shallow minimum and the
rank-one saddle that merge along that single soft direction \cite{Aspelmeier04, Cavagna2004,Mueller06}.

A different notion, adopted in Refs.~\cite{KentDobias24,KentDobias26}, calls
marginal those stationary points whose Hessian spectral density is
\emph{pseudogapped}, i.e.\ vanishes continuously at the origin, so that a
finite density of arbitrarily soft modes is present. In the spherical models
such points are generically \emph{subdominant}, and are selected by
conditioning the count on the value of the smallest eigenvalue, or on the trace
of the inverse Hessian. The two definitions might not coincide in the SK model,
in which the marginality of the typical states is carried by a single isolated
null eigenvalue \cite{Aspelmeier04,Parisi04b}, a rank-one property
contributing vanishing weight to the spectral density, which therefore cannot
by itself produce (or hinder) a pseudogap: the two notions are predicates on
different objects. Refs.~\cite{KentDobias24,KentDobias26} state that the
typical stationary points of the SK model are not marginal in the pseudogap
sense. Within the annealed approximation this is corroborated by the Plefka
parameter, $x_P>0$ with finite $\chi_{SG}$ along the whole physical branch
(Appendix~\ref{subsec:stability}), which excludes a spectral density vanishing at the
origin as $\lambda^{\alpha}$ with $\alpha\le1$. Indeed for $|x_P|\ll 1$,
$\lambda \ll 1$, $N\gg 1$, the bulk eigenvalue distribution for the annealed
solution is known to behave as \cite{Crisanti03c,Plefka02}
\begin{equation}
 \rho(\lambda) \propto \sqrt{\lambda-\frac{x_P^2}{4\delta_3}}\,,
 \qquad
 \delta_3\equiv \frac{1}{N}\sum_{i=1}^N\left[\beta(1-m_i^2)\right]^3.
\end{equation}

What is the quenched expression of $x_P$ and whether it stays strictly
positive along the quenched branch below $f^{*}$ has not been determined yet.

A third independent signature of marginality is given by the distribution of local fields $P(h)$, which is the analogue for spin glasses of the single-particle
density of states of electron glasses and of the contact-force distribution of packings \cite{Anderson1979,Palmer1979,MuellerWyart15}. Stability against the flip of a finite number of spins requires $P(h)$ to be suppressed linearly at the origin in marginally stable states. In the SK model, this linear pseudogap, $P(h)\to c\,|h|$ as $h\to0$ with $c\simeq0.31$, is a property of the full-RSB \emph{equilibrium} 
state~\cite{SommersDupont84,Pankov06}, equivalent to the vanishing of the replicon through the marginality condition.

\begin{figure}
    \centering
    \includegraphics[width=\linewidth]{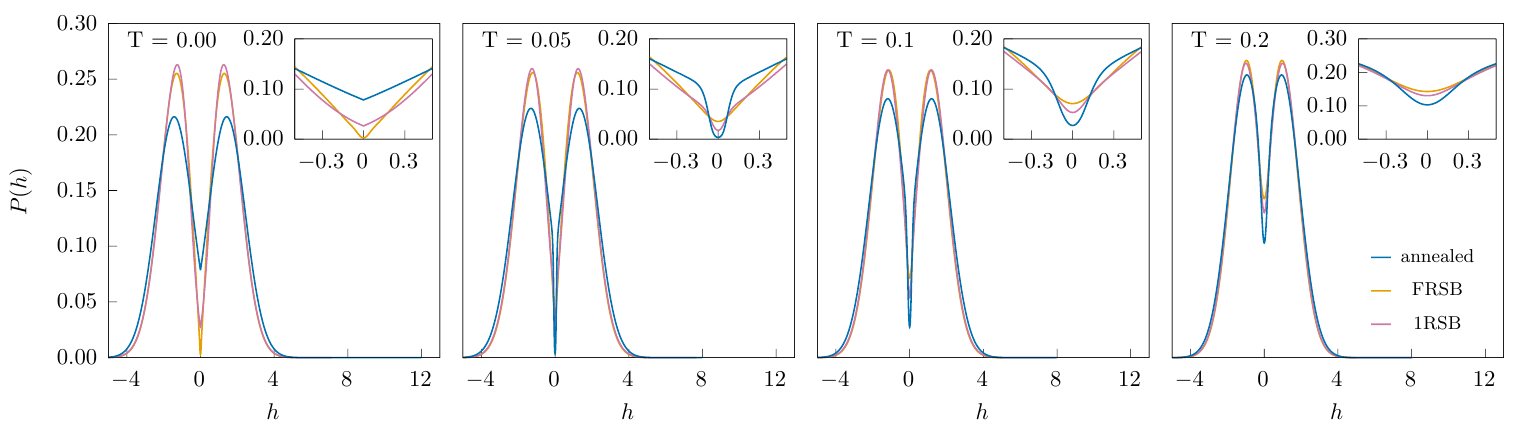}
    \caption{Distribution of local fields $P(h)$ in the annealed (blue), quenched 1RSB (pink) and full-RSB (yellow) ansatz, evaluated respectively for their zero-complexity point, and for different values of temperature. The distributions are all normalized to unity.}
    \label{fig:entropy_allT}
\end{figure}

Fig. (\ref{fig:entropy_allT}) shows a comparison among the field distribution $P(h)$, evaluated in the zero-complexity point, for the annealed approximation, the 1RSB quenched approximation, which are both unstable in that point, and finally the quenched stable FRSB ansatz. For finite temperatures the difference between the approximations and the exact FRSB solution are less evident, whereas at $T=0$ only the FRSB equilibrium state develops the expected pseudogap $P(0)=0$.

\clearpage
\bibliography{Lucabib}

\end{document}